\documentclass[11pt]{article}

\usepackage{amsmath,amssymb,exscale}
\usepackage{array,multicol}
\usepackage{afterpage,float,flafter}
\usepackage{amssymb}
\usepackage{epsfig,rotating,pifont}
\usepackage{cite}
\usepackage{textcomp}
\@addtoreset{equation}{section} \makeatother

\def\beq{\begin{eqnarray}}
\def\eeq{\end{eqnarray}}
\def\lsim{\mathrel{\rlap{\lower3pt\hbox{\hskip0pt$\sim$}}
\raise1pt\hbox{$<$}}}         
\def\gsim{\mathrel{\rlap{\lower4pt\hbox{\hskip1pt$\sim$}}
\raise1pt\hbox{$>$}}}         

\def\eg{{\it e.g.}}
\def\vs{{\it vs}}
\title{
\vspace{-3cm}
\begin{flushright}
\small{CERN-PH-TH/2010-202}\\
\small{UCSD PTH/10-07}
\end{flushright}
\vspace{2.7cm}
\huge{Low Scale Flavor Gauge Symmetries}
\vspace*{0.7cm}
\author{
\Large{\text{Benjam\'\i{}n Grinstein$^{1,2}$}\footnote{bgrinstein@ucsd.edu},~~~\text{Michele Redi$^1$}\footnote{michele.redi@cern.ch}~~{\large and} \text{Giovanni Villadoro$^1$}\footnote{giovanni.villadoro@cern.ch}}\\ \\
$^1$\emph{CERN, PH-TH, CH-1211, Geneva 23, Switzerland}\\
$^2$\emph{Physics Dept., UCSD, La Jolla, CA 92093, USA}
}
}

\date{}
\begin{document}
\maketitle \thispagestyle{empty} \vspace*{-.2cm}

\begin{abstract}
We study the possibility of gauging the Standard Model flavor group. Anomaly cancellation 
leads to the addition of fermions whose mass is inversely proportional to the known fermion masses. 
In this case all flavor violating effects turn out to be controlled roughly
by the Standard Model Yukawa, suppressing transitions for the light generations.
Due to the inverted hierarchy the scale of new gauge flavor bosons could be as low as the electroweak scale 
without violating any existing bound but accessible at the Tevatron
and the LHC. 
The mechanism of flavor protection potentially provides an alternative to Minimal Flavor Violation, 
with flavor violating effects suppressed by hierarchy of scales rather than couplings. 

\end{abstract}

\newpage
\renewcommand{\thepage}{\arabic{page}}
\setcounter{page}{1}

\section{Introduction}

In the Standard Model (SM) the only source of flavor violation arises
from the Yukawa couplings. In the limit of vanishing quark masses the
SM Lagrangian acquires a large global symmetry (known as flavor or
horizontal symmetry) mixing SM fermions of different generations.  It
is quite tempting to impose such a symmetry on the physics beyond the
SM in order to suppress extra flavor violation. In the
extreme case where all new physics effects are flavor universal up to
small corrections satisfying the full flavor symmetry and
proportional to the SM Yukawa, the scale of such new physics can be a
TeV. This idea, known as minimal flavor violation (MFV) is quite
old \cite{Chivukula:1987py,Hall:1990ac}, but recently is gaining ever more
interest (see \cite{D'Ambrosio:2002ex,Cirigliano:2005ck}) as
a consequence of the persistent failure to find flavor violation
beyond the Standard Model.

The idea of assuming horizontal symmetries to be true symmetries of
nature is even older \cite{Barr:1978rv}.  Unfortunately such
assumption itself is not enough to suppress flavor violation below
the experimental bounds when the flavor symmetry is broken at low
scales.  The classical arguments against low-scale flavor symmetry
work as follows (see, \eg,  \cite{ArkaniHamed:1999yy} for a nice
review).  In order to produce the SM Yukawa couplings flavor symmetry must
eventually be broken by the vacuum expectation value (VEV) of some field
(``flavon'').  This implies the presence of
massless Goldstone bosons (GB) and bounds from hadron decays and
astrophysics on such states are even stronger than those from flavor
physics. Of course GBs can be easily avoided by
gauging. In this case, however, there are flavor gauge bosons that
mediate dangerous flavor changing neutral currents (FCNC) and their
masses must be well above the TeV scale.\footnote{Gauging only an
abelian subgroup would not help because FCNC are reintroduced after
the rotation needed to go to the mass eigenstate basis.}

Even requiring that the only sources of flavor breaking are the SM
Yukawas is not enough to avoid large FCNC from the flavor gauge
bosons. Indeed if the masses of the gauge bosons are proportional to
the SM Yukawa couplings, they generate tree-level four-fermion
operators proportional to inverse powers of the SM Yukawa couplings,
enhancing FCNC among the first generations. Therefore despite the fact
that the only spurions breaking flavor are the SM Yukawa couplings,
as in MFV, inverse powers of the spurions appear in the higher
dimensional operators, producing \emph{de facto} ``maximal'' flavor
violating operators.  This argument shows how MFV models cannot arise
directly from having a fundamental flavor symmetry in the underling
theory but rather from accidental ones.

In the classical argument against low-scale flavor gauge bosons
sketched above we can easily identify a way out.  If the fields
breaking the flavor symmetry are instead proportional to the inverse
of the SM Yukawa couplings, the effective operators generated by integrating out
the flavor gauge bosons will be roughly proportional to positive
powers of the Yukawa couplings, suppressing flavor violating effects
for the light generations, much like in MFV models.  The spectrum of
the extra flavor states, controlled by the flavon VEVs, will thus present an
inverted hierarchy, with states associated to the third generation
much lighter than those associated to the first two. 
Models implementing this inverted hierarchy were first introduced 
in \cite{Berezhiani:1983rk,Berezhiani:1990wn}.

Remarkably, as we will show in this paper, the mechanism described above
is automatic in the minimal extension of the SM with gauged flavor symmetries.
With just the SM fermion content the full SM plus
flavor gauge group would be anomalous. Extra flavorful fermions have
to be added to cancel these anomalies.  Such fermions are also welcome
as they can make the SM Yukawa terms arise from a renormalizable
Lagrangian, now that the Yukawa couplings have been uplifted to
dynamical ``flavon'' fields. The smallest set of fermions canceling all anomalies leads
automatically to the inverted-hierarchy structure mentioned
above. The quantum numbers of these extra fermions are indeed such
that the mixing with the SM fermion is flavor diagonal while their
masses are proportional to the flavon VEVs. The SM fermion masses arise
via a see-saw like mechanism, after integrating out the extra
fermions, and are thus proportional to the inverse of the flavon
VEVs.  All non-SM particle masses (fermions, vector bosons and flavon
fields) are controlled by the flavon VEVs, thus they are roughly
proportional to the inverse of the SM Yukawa.  The resulting inverted
hierarchy in the new physics sector protects the SM fermions from
getting large flavor breaking effects even when the lightest new
states lie at the electroweak scale.

There are a number of analogies with MFV models:
new physics effects are controlled by the flavor group, we may
have models where only one spurion for each SM Yukawa matrix 
breaks the flavor symmetry, and the flavor breaking
effects follow the hierarchical structure of the Yukawa couplings.  
However, this kind of models are not MFV.  Indeed,
in these models there is a limit where all Yukawa
couplings vanish but flavor breaking effects remain finite.  
Contrary to the naive intuition that flavor violating effects must be
larger than in MFV,
it is very easy to find values of the parameters that produce extra
flavor non-universal states without incurring into a flavor
problem. Moreover, unlike in MFV, these could 
be light, even below the electroweak scale.  
The tightest bounds on this kind of models do not come from flavor breaking observables
but rather from electroweak precision tests (EWPT) and direct searches
for new particles, opening the possibility for direct discoveries of
flavor physics at the LHC.

The mechanism protecting from flavor violations is robust
against deformations of the model, both when more flavon fields are
considered and when the nature of the flavor subgroup that is
actually gauged is changed.  Of course the detailed structure of the
flavor sector as well as the size of the flavor violations will
depend on these modifications but the latter will continue to remain
sufficiently small in most of the parameter space of the theory.

We should point out that the possibility of having non-universal gauge
bosons (and other flavorful physics) at the TeV scale is well known
in the literature. In composite Higgs models and similar
extra-dimensional constructions, for example, there is the possibility
of having extra flavorful gauge bosons. Unlike in our case however,
these vector fields are not ``the'' flavor gauge bosons (\emph{viz.}
they are not the states eating the Goldstone bosons of the broken
flavor symmetry), but rather gauge bosons in some non-trivial
representation of the flavor group, getting mass splitting from the
breaking of the flavor symmetry (which generically is explicit in
these models).  In this respect they are closer to realizing the MFV
idea~(see, \eg, \cite{Rattazzi:2000hs,Cacciapaglia:2007fw}).

In the rest of the paper we give the details on how the mechanism works,
we will discuss where the strongest bounds on the model come from and
possible signatures at hadron colliders. For definiteness we will
focus on the quark sector, gauging the full flavor group and
considering mainly the minimal set of flavon fields, although the
same mechanism can easily be applied to more general situations.

\section{Inverted Hierarchies From Anomaly Cancellation}
\label{sec:InvHi}

In the absence of Yukawas, focusing on the quark sector,  the SM enjoys at the classical level the
global symmetry
\begin{equation}
U(3)_{Q_L}\otimes U(3)_{U_R} \otimes U(3)_{D_R}\,,
\label{flavorsymmetry}
\end{equation} 
where $Q_L$, $U_R$ and $D_R$ transform as fundamentals.

We assume this to be an exact symmetry of nature. In order to allow Yukawa couplings
the flavor symmetry should be broken spontaneously by the vacuum. 
This can be most simply realized by the VEVs of two bifundamentals flavon fields
transforming as
\begin{equation}
\begin{aligned}
Y_u&=(\bar{3},3,1)\,, \\
Y_d&=(\bar{3},1,3)\,.
\end{aligned}
\end{equation}
In general the VEVs of these fields, while related, should not
be confused with the Yukawa matrices, as functions of $Y_{u,d}$ may have equal
transformation properties. Indeed this will be the crucial feature of
our model. To avoid problematic flavor violating GBs, the symmetry should be gauged.  
Within the SM the gauging of the SM flavor symmetry (\ref{flavorsymmetry}) is anomalous
due to cubic and mixed hypercharge anomalies.  The simplest option 
to cancel the cubic non-abelian anomalies is to add two right-handed colored fermions in the
fundamental of $SU(3)_{Q_L}$, one left handed fundamental of
$SU(3)_{U_R}$ and one left-handed fundamental of $SU(3)_{D_R}$. In this way the fermions 
are vector-like with respect to the flavor gauge group but remain chiral with respect to the SM gauge symmetry.
The other possibility, with the two right-handed triplets in an
$SU(2)_L$ doublet is an uninteresting, non-chiral model. We are
therefore led rather uniquely to the following model:

\begin{center}
\begin{tabular}{c|cccccc}
				& 	SU(3)$_{Q_L}$ 	&	SU(3)$_{U_R}$ 	&	SU(3)$_{D_R}$	&	SU(3)$_c$	&SU(2)$_L$	& 	U(1)$_Y$\\ \hline
$Q_L$				&	$3$			&	1			&	1			&	3			&	2	&	1/6		\\
$U_R$			&		1		&	$3$			&	1			&	$3$			&	1	&	2/3		\\
$D_R$			&		1		&	1			&	$3$			&	$3$			&	1	&	-1/3	\\
$\Psi_{uR}$		&	$3$			&	1			&	1			&	$3$			&	1	&	2/3		\\
$\Psi_{dR}$		&	$3$			&	1			&	1			&	$3$			&	1	&	-1/3	\\
$\Psi_{u}$		&	1			&	3			&	1			&	3			&	1	&	2/3		\\
$\Psi_{d}$		&	1			&	1			&	3			&	3			&	1	&	-1/3	\\
$Y_u$			&	$\overline3$&	$3$			&	1			&	1			&	1	&	0		\\
$Y_d$			&	$\overline3$&	1			&	$3$			&	1			&	1	&	0		\\
$H$				&	1			&	1			&	1			&	1			&	2	&	1/2
\end{tabular}
\end{center}
Remarkably, with the above matter content all the anomalies except $U(1)_{Q_L} \times SU(2)_L^2$ and $U(1)_{Q_L}
\times U(1)_Y^2$ automatically cancel. When, as required by cancellation of SM anomalies, the leptons are introduced
$U(1)_{B-L}$ remains anomaly free, so that $U(1)_{Q_L}$ could also be
gauged by gauging the $B-L$ combination.
The VEVs of $Y_u$ and $Y_d$ break $U(1)_{Q_L}\times
U(1)_{U_R}\times U(1)_{D_R}$ to the diagonal $U(1)$ and an additional scalar
field must be introduced in order to break also $U(1)_{B-L}$  spontaneously.  
From now on we will focus on the gauging of $SU(3)^3\times U(1)^2$ which is the largest symmetry group 
broken by the SM Yukawa, other gaugings will be considered later.

The most general renormalizable Lagrangian reads,
\begin{equation}
\begin{aligned}
{\cal L}=& {\cal L}_{kin}-V(Y_u,Y_d,H)+\\
&\bigl( \lambda_u\, \overline Q_L \tilde H \Psi_{uR}+\lambda_u'\,\overline \Psi_{u} Y_u \Psi_{uR} + M_u\,  \overline \Psi_{u} U_R+ \\
&\lambda_d\, \overline Q_L H \Psi_{dR}+ \lambda_d' \,\overline \Psi_{d} Y_d \Psi_{dR} + M_d\,  \overline \Psi_{d} D_R+h.c.\bigr)\,,
\label{action}
\end{aligned}
\end{equation}
where $M_{u,d}$ are universal mass parameters and
$\lambda^{(\prime)}_{u,d}$ are universal coupling constants. By a rotation of $\Psi_u$ and $\Psi_{uR}$ these 
parameters can be chosen to be real. 
The kinetic terms are built from covariant derivatives, which in our conventions are given by
\begin{equation}
{\cal D} Q_L=\partial Q_L+i g_Q A_Q Q_L+i g_3 A_c Q_L+i g W Q_L +ig'{\textstyle\frac16} B Q_L 
\end{equation}
and similarly for the other fields.

In general, the VEVs of $Y_{u,d}$  break the flavor symmetry to baryon
number.\footnote{We use the same notation both for the fields
 $Y_{u,d}$ and their VEVs, except when the meaning is not immediate
 from the context.} By a flavor transformation we can take
$Y_d=\hat{Y}_d$ diagonal and $Y_u=\hat{Y}_u V$ where $V$ is a unitary
matrix. Integrating out the heavy fermions generates Yukawa
interactions for the SM fields.  At leading order for
$Y_{u,d}\gg M_{u,d}$ one immediately finds that the Yukawa couplings of the
SM are
\begin{equation}
\begin{aligned}
\label{yukawas}
y_u&=  V^\dagger \frac {\lambda_u M_u}{\lambda_u'\hat{Y}_u}\,,\\
y_d&=  \frac {\lambda_d M_d}{\lambda_d'\hat{Y}_d}\,.
\end{aligned}
\end{equation}

Importantly the masses of the SM fermions follow an inverted hierarchy
controlled by the inverse of $\hat{Y}_{u,d}$ (see also 
\cite{Berezhiani:1983rk,Berezhiani:1990wn}  for related works
implementing the inverted hierarchy mechanism 
with models where the chiral diagonal SU(3) flavor symmetry is gauged). On the other hand, 
the exotic fermions have a mass proportional to $\hat{Y}_{u,d}$ so that the lightest partner is the one 
associated to the top quark. As we will see this kind of see-saw mechanism
is a general feature of the model through which all flavor and electroweak
precision bounds can be easily avoided. The unitary matrix $V$ plays the
role of the CKM matrix of the SM. The formulas above receive important
corrections for the third family since in this case the condition $Y_{u,d}\gg M_{u,d}$ is
not satisfied, particularly for the top quark.  As we will see in the next
section once this is properly accounted for it modifies the SM couplings. This produces 
important corrections to precision observables, in particular to the electroweak oblique parameters and  the
$Zb\bar{b}$  coupling, which impose the most stringent bounds on the model.

\subsection{Vectors and Scalars}
\label{sec:VandS}

The VEVs of $Y_{u,d}$ give also a mass to the flavor gauge bosons,
\begin{align}
\label{eq:Lmass}
{\cal L}_{mass}
&= {\rm Tr} |g_U A_U Y_u-g_Q Y_u A_Q|^2 + {\rm Tr}  |g_D A_D Y_d-g_Q Y_d A_Q |^2 \nonumber \\
&=\frac12 V_{Aa} (M_V^2)^{Aa,Bb} V_{Bb} \,,
\end{align}
where
\begin{equation}
V_{Aa}=\left\{A_{Q\,a}\,,A_{U\,a}\,,A_{D\,a}\right\}\,,\quad
 A_{Q}=A_{Q\,a}\,\frac{\lambda^a}{2}\,,\quad A_{U}=A_{U\,a}\,\frac{\lambda^a}{2}\,,\quad A_{D}=A_{D\,a}\,\frac{\lambda^a}{2}\,,
\end{equation}
$\lambda^{a=1,\ldots,8}$ are the Gell-Mann matrices and $\lambda^9$ is proportional to the identity. 

The flavor gauge bosons couple to the quark currents,
\begin{equation}
J^{\mu\,ij,A} = (g_Q {\overline Q}_L^i \gamma^\mu  Q_L^j,\, 
g_U{\overline U}^i_R \gamma^\mu  U_R^j ,\, 
g_D{\overline D}^i_R \gamma^\mu D_R^j ).
\end{equation}
Integrating out the vector fields SM four-fermion operators are produced, which in the flavor basis read
\begin{equation} \label{eq:4fermop}
-\frac18 (M_V^2)^{-1}_{Aa,Bb}\, \lambda^{a}_{ij}\lambda^{b}_{hk}\, J_\mu^{ij,A}\  J^{\mu\,hk,B}\,.
\end{equation}
In order to get the four-fermion operators in the mass eigenstate basis a further rotation by the unitary matrix 
$V$ is needed on the left-handed up-quarks.

The flavor gauge bosons mediate FCNC since their masses break all flavor symmetries. 
Naively this implies the masses of all the gauge bosons to be around $10^5$~TeV or higher in order to comply with
flavor bounds. This expectation is however completely incorrect in our
model because the masses depend on the inverse Yukawas. Roughly
speaking the gauge bosons associated with transitions between light
generation are automatically much heavier than the ones associated with 
the third generation with a hierarchy determined by the inverse
Yukawas. As a consequence FCNC, which roughly scale as
\begin{equation} \label{eq:rough4q}
\sim \frac{1}{Y_{u,d}^2}(\bar q\gamma^\mu q)^2\,,
\end{equation}
are highly suppressed for the light generations.

To better understand how this works let us consider for simplicity the case where only $Y_u$
is present. Since $Y_u$ can be taken to a diagonal form there are no
flavor violating processes and the individual family numbers are not
broken so the associated gauge bosons remain massless. The masses of
the flavor gauge bosons can be computed analytically in this
case. Assuming equal couplings for $SU(3)_{Q_L}$ and $SU(3)_{U_R}$ the
mass terms can be written as follows,
\begin{align}
{\cal L}_{mass}&=\frac12 g^2 |V_{ij}|^2 (\hat{Y}_u^{i} -\hat{Y}_u^{j} )^2 
+\frac12 g^2  |A_{ij}|^2 (\hat{Y}_u^{i} + \hat{Y}_u^j )^2 \nonumber \\
&\approx \frac12 g^2 |V_{ij}|^2 \left(\frac{\lambda_u M_u}{\lambda'_u}\right)^2 
\left(\frac{1}{y_u^i}-\frac{1}{y_u^j}\right)^2 
+\frac12 g^2  |A_{ij}|^2 \left(\frac{\lambda_u M_u}{\lambda'_u}\right)^2 
\left(\frac{1}{y_u^i}+\frac{1}{y_u^j}\right)^2\,,
\end{align}
where $V$ and $A$ are the combinations $(A_Q+A_U)/\sqrt2$ and $(A_Q-A_U)/\sqrt2$
respectively.  From this it follows that 4-fermion operators with light quarks obtained
integrating out heavy gauge bosons are very suppressed.  The same
mechanism works once the effects of $Y_d$ are included, where all the
flavor symmetries are broken and FCNC are generated.  As we will show
in various examples, flavor constraints can be avoided generically
even if the lightest gauge boson is below the TeV scale.

The scalar sector is more model dependent due to the unspecified scalar potential. We discuss the radial 
fluctuations in detail in  appendix \ref{appendixscalar}.
After flavor symmetry breaking there are 10 radial fields contained in $Y_{u,d}$, corresponding 
to fluctuations of quark masses and CKM angles. These modes couple to fermion bilinears and therefore
generate at low energy four-fermion operators. In particular fluctuations of the masses give rise to flavor diagonal operators
and fluctuations of the CKM matrix induce flavor changing processes. However the suppression due to the inverted 
hierarchy works in this sector as well.  To get an intuition for why this is the case we focus again on the 
flavor preserving four-fermion operators induced by the mass fluctuations.
Their coupling to quarks is given by (for values of the couplings $\lambda_{u,d}$ and $\lambda'_{u,d}$ of order one)
\begin{equation}
\sim \frac M {\hat{Y}^i+\delta Y^i} v \,\bar{q}_i q_i \,\approx \left ( 1 -y^i\frac{\delta Y^i}{M} \right )m_{q_i} \bar{q}_i q_i \,,
\end{equation}
so that the couplings of the radial modes $\delta Y_i$ are highly suppressed for the light generations.
Since these modes unitarize the scattering of the massive gauge bosons
we expect their masses to be naturally set by the VEVs  $(m_{\delta Y_i}\sim \hat{Y}_i)$.
In this case the coefficients of the four-fermion operators scale as
\begin{equation}
\sim \left(\frac{y^i m_{q_i}}{M \hat Y^i}\right)^2
\end{equation}
which is extremely suppressed for the light generations. 
Actually  the highly suppressed couplings alone would be enough to suppress dangerous
four-fermion operators even when the flavon fields are light.

\subsection{Remarks}

A few comments are in order. 
While in MFV in the limit of vanishing Yukawa couplings the full flavor symmetry is restored,
in our model there exists a limit where all Yukawa couplings vanish but flavor-breaking contributions
remain finite. This can be seen by sending $M_{u,d}\to 0$ with all other parameters fixed.
In this limit $y_{u,d}\to 0$ (see Eq.~(\ref{yukawas})) while four-fermion operators,
depending only on $Y_{u,d}$ (see Eq.~(\ref{eq:rough4q})), still break flavor.

In the model above we assumed for simplicity the existence of only two bifundamentals.
Actually in this case it can be shown that there is no renormalizable potential that gives rise to 
the Yukawa pattern of the SM. One possibility is to introduce non-renormalizable potentials. As long as
the cut-off suppressing higher-dimensional operators is larger than the largest flavon VEV, its effects
can be treated as perturbations, without spoiling our mechanism.
Alternatively the $Y_{u,d}$ could be combinations of several fields transforming 
as bifundamentals under the flavor group,
\begin{equation}
Y_{u,d}=\sum_{i=1}^N a_{u,d}^i X_{u,d}^i\,.
\end{equation}
We have checked that in this case models with renormalizable potentials can be build.
The mechanism of inverted hierarchy is still at work, leaving the fermion sector as before, 
however the relation between the Yukawas  and the gauge boson masses is not uniquely determined. 
Unless the VEVs of the different fields are correlated the flavor gauge bosons will be generically
heavier than in the minimal case improving flavor bounds but limiting the possibility of having these states 
at the electroweak scale. 

In this paper we focus on the flavor symmetries of the quark sector but it is straightforward to extend this analysis 
to leptons at least when right-handed neutrinos are included. In this case the SM flavor symmetry is $U(3)^6$. 
For leptons cancellation of cubic anomalies works similarly to the quark sector and requires
the addition of fermions transforming as singlets of $SU(2)_L$ and with hypercharge opposite to the SM. In this way one finds that the only 
anomalous flavor symmetry is $U(1)_{B+L}$. We leave the detailed investigation of the lepton sector to future work. 

One could also consider the  gauging of smaller subgroups of the SM flavor symmetry. 
Obviously cancellation of anomalies can be achieved with the same matter content considered here so that 
the  mechanism of inverted hierarchy works as before. An interesting subgroup is the diagonal 
$SU(3)$ subgroup where the SM left- and right-handed fermions transform as fundamentals and anti-fundamentals respectively.\footnote{The other choice where the left and right fermions are fundamentals is already anomaly free within the SM and has been considered in the past, see for example \cite{kribs} and Refs. therein.} In this case however the mass of the $SU(3)$  flavor gauge bosons is necessarily increased. 
Another interesting example is the gauging of abelian subgroups as also in this case, due to the inverted hierarchy,
large corrections to FCNC do not arise. 

Concerning unification the addition of the new fields charged under color and hypercharge 
worsens the unification of gauge couplings in the SM.  Moreover in the case of $SO(10)$ 
unification the flavor symmetry is only $SU(3)$. The simplest way to cancel the flavor cubic anomaly is
to add fermions in the anti-fundamental representation of the
flavor symmetry and the {\bf 16} of $SO(10)$ to leave the theory chiral.
However these degrees of freedom are insufficient to generalize our model since only one Yukawa term can be written down. 
Also for $SU(5)$ the inverted hierarchy structure cannot
be obtained at least in the simplest constructions. It is unclear to us how  this model could be embedded in unification.

\section{Experimental Bounds}

\label{sec:modsmcoup}

In this section we consider the experimental bounds arising from the exotic fermions. 
These are the most model independent limits on the model as they only depend on four parameters,
which can be conveniently chosen as the ratios $\lambda_{u}/y_{t}$,  $\lambda_{d}/y_{b}$,
 $M_{u}/m_{t}$ and $M_{d}/m_{b}$. The bounds originate
from mixing effects between SM and exotic fermions that contribute in particular to $Z\to b_L \bar{b}_L$, 
EW oblique parameters, $b\to s \gamma$ and $V_{tb}$ as well as  from direct searches.

In the previous section we integrated out the exotic fermions to
leading order assuming $Y_{u,d}\gg M_{u,d}$. This is in general
insufficient for the third family, in particular for the top whose
large Yukawa requires a large or maximal mixing between SM and exotic
fields and for the bottom whose coupling to $Z$ may receive observable corrections.

The fermion mass matrices can be easily diagonalized in general. We can first eliminate the matrix
$V$ from the Yukawa interactions in eq. (\ref{action}) by a simultaneous rotation of $u_L$ and $\Psi_{uR}$. The
physical states (without renaming the fields) are then given by the orthogonal rotations of the left and right fields, 
\begin{eqnarray}
\left ( \begin{array}{c} u^i_{R} \\ u'^i_{R} \end{array} \right)
= \left ( \begin{array}{cc} c_{u_{Ri}} & -s_{u_{Ri}} \\ s_{u_{Ri}} & c_{u_{Ri}} \end{array}\right) \left(\begin{array}{c} U^i_{R}\\ \Psi^i_{uR} \end{array} \right)
\,, \qquad
\left ( \begin{array}{c} u_{L}^i \\ u'^i_{L} \end{array} \right)
= \left ( \begin{array}{cc} c_{u_{Li}} & -s_{u_{Li}} \\ s_{u_{Li}} & c_{u_{Li}} \end{array}\right) \left(\begin{array}{c} U^i_{L}\\ \Psi^i_{u} \end{array} \right)\,,
\end{eqnarray}
and similarly for the down-quark sector. The masses of SM and heavy fermions are then given by,
\begin{equation}
\begin{aligned}
(m_u,m_c,m_t)&= \lambda_u \frac v {\sqrt{2}}\,
	\left(\frac {s_{u_{R1}}}{c_{u_{L1}}},\frac {s_{u_{R2}}}{c_{u_{L2}}},\frac {s_{u_{L3}}}{c_{u_{L3}}}\right)\,, \\
(m_{u'},m_{c'},m_{t'})&=  M_u\,
	\left(\frac {c_{u_{L1}}}{s_{u_{R1}}},\frac {c_{u_{L2}}}{s_{u_{R2}}},\frac {c_{u_{L3}}}{s_{u_{L3}}}\right)\,.
\end{aligned}
\end{equation}
We find it useful to define the physical variables,
\begin{equation}
\label{eq:defxyetc}
x_i \equiv \frac{M_u}{m_{u_i}}\,,~~~~~~~~~~\qquad y_i\equiv\frac{\lambda_u v}{\sqrt2 m_{u_i}}\,,
\end{equation}
which satisfy the properties,
\begin{equation}
\label{eq:defxyprop}
x_i = \frac{c_{u^i_{R}}}{s_{u^i_L}}\,,\qquad y_i=\frac{c_{u^i_{L}}}{s_{u^i_R}}\,,\qquad
\frac{m_{u'_i}}{m_{u_i}}=x_i y_i\,,\qquad \frac{\lambda'_u \hat Y^i_u}{m_{u_i}}=\sqrt{(x_i^2-1)(y_i^2-1)}\ .
\end{equation}
From the above relations one can easily derive,
\begin{align}
\label{suL}
s_{u_{Li}} &=\sqrt{\frac{y_i^2-1}{x_i^2 y_i^2-1}}= \frac{\lambda'_u \hat Y^i_u
  m_{u_i}}{\sqrt{(M_u^2-m_{u_i}^2)^2+(\lambda'_u \hat Y^i_u
    M_u)^2}}
 \,,\\
\label{suR}
s_{u_{Ri}} &=\sqrt{\frac{x_i^2-1}{x_i^2 y_i^2-1}} = \frac{\lambda'_u \hat Y^i_u
  m_{u_i}}{\sqrt{(\frac12(\lambda_uv)^2-m_{u_i}^2)^2+\frac12(\lambda_u\lambda'_u \hat Y^i_u v)^2}}
\,.
\end{align}
Note that the physical region of $x_i$ and $y_i$ corresponds to
$x_i,y_i\ge1$ or  $x_i, y_i<1$. 
In the first case $m_{u'_i}\ge m_{u_i}$ while  $m_{u'_i}\le m_{u_i}$  in the second. 
In the limit $y_3 \to 1$ ($\lambda_u\to y_t=\sqrt2 m_t/v$), corresponding to $\hat{Y}_u^3\to 0$,
the right handed top becomes $\Psi_{uR}^3$ while $U_R^3$ becomes the right handed top-prime.

For phenomenological purposes and to better understand the parametric
dependence of the results the following approximate expressions will
be  useful too:
\begin{equation}
\label{slsr}
\begin{aligned}
s_{u_{Li}}&=\frac{\lambda_u \lambda'_u\, v \hat{Y}_u^{i}}{\sqrt2
  (M_u^2+{\lambda'_u}^2 \hat{Y}_u^{i2})} \,, \\
s_{u_{Ri}}&=\frac{M_u}{\sqrt{M_u^2+{\lambda'_u}^2 \hat{Y}_u^{i2}}}
	\left (1-\frac{\lambda_u^2 {\lambda'_u}^2 v^2 \hat Y_u^{i2}}{2 (M_u^2+{\lambda'_u}^2 \hat{Y}_u^{i2})^2} \right )\,,
\end{aligned}
\end{equation}
valid up to terms
${\cal O}(v^3)$ in the expansion in the SM Higgs VEV with respect to
the new scales ($\hat Y_t$ and $M_u$). Note that before electroweak symmetry breaking only the right-handed quarks
mix with the exotic fermions $\Psi_{u,dR}$ while the left-handed mixing is suppressed by the Higgs VEV. 

The most important  consequence of the mixing is that the quark couplings are modified relative to
those in the SM. For example the charged current (that couples to
$g_2W^+_\mu/\sqrt2$) in terms of mass eigenstates is 
\begin{equation}
\bar u_L (c_{u_L} V c_{d_L})\gamma^\mu d_L
+\bar u_L (c_{u_L} Vs_{d_L})\gamma^\mu d'_L
+\bar u'_L (s_{u_L} V c_{d_L})\gamma^\mu d_L
+\bar u'_L (s_{u_L} V s_{d_L})\gamma^\mu d'_L.
\label{Wcouplings}
\end{equation}
We have used the shorthands $c_{u_L}=\text{Diag}(c_{u_{L1}},
c_{u_{L2}}, c_{u_{L3}})$, etc, and $V$ is the unitary matrix
introduced in (\ref{yukawas}). Effectively the CKM matrix now becomes
\begin{equation}
\label{CKMeff}
V_{CKM}= c_{u_{L}}\cdot V\cdot c_{d_{L}}\,.
\end{equation}
Note that such a matrix is not unitary. However, as we will see
shortly, all the $s_{q_{Li}}$ are exceedingly small except, possibly,
for that of the top quark. Moreover, the $6\times6$ matrix of couplings to the
charge current {\it is} unitary, hence exhibiting a generalized GIM
mechanism. 

The couplings of quarks to the photon are not modified, since they are
protected by gauge invariance. And since the right handed quarks only mix with singlets of equal charge
their couplings to the $Z$ (proportional to their electric charge)
are not modified either. The coupling of left handed quarks to the $Z$ is  now
through the current
\begin{equation}
\label{Zcouplings}
\bar u_L (T_3^{u} c_{u_L}^2-s_w^2Q_u)\gamma^\mu u_L
+\bar u_L(T_3^u c_{u_L}s_{u_L})\gamma^\mu u'_L
+\bar u'_L(T_3^u s_{u_L}c_{u_L})\gamma^\mu u_L
+\bar u'_L(T_3^u s_{u_L}^2-s_w^2Q_u) \gamma^\mu u'_L +(u\to d),
\end{equation}
where $Q_{u(d)}=2/3(-1/3)$ and $s_w$ is the sine of the weak mixing
angle. Using Eq.~(\ref{slsr}) we see that $\delta g_{b_L}/g_{b_L}\sim (m_b/M_d)^2$. 
The couplings of quarks to the Higgs are also modified relative
to those in the SM:
\begin{equation}
\label{hcouplings}
{\textstyle\frac1{\sqrt2}}\lambda_uh[-\bar t_Lc_{u_L}s_{u_R}t_R
+\bar t_Lc_{u_L}c_{u_R}t'_R-\bar t'_Ls_{u_L}s_{u_R}t_R+\bar t'_Ls_{u_L}c_{u_R}t'_R]+(u\to d)+\text{h.c.}
\end{equation}

\subsection{Bounds from the down sector}

\begin{figure}[t!] 
    \centering
    \includegraphics[width=\textwidth]{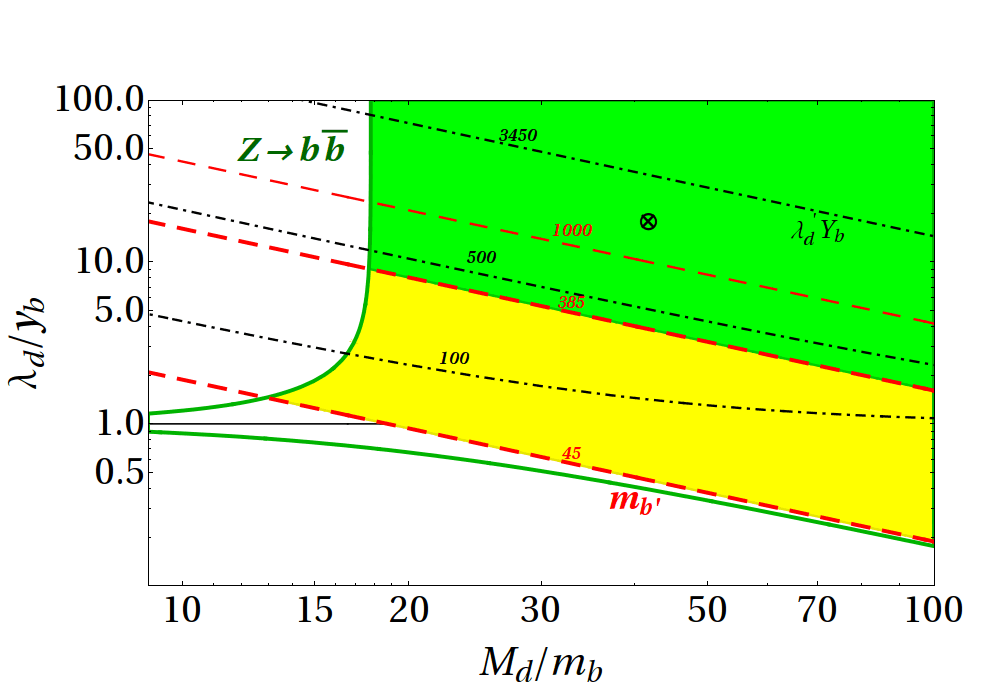}
    \caption{Allowed region of parameter space in the $\lambda_d$ vs
      $M_d$ plane. The yellow and green shaded regions  are
      allowed by $R_b$, the thick green line labeled $Z\to b \bar b$
      corresponding to the 95\% CL limit.  The green one corresponds
      to $m_{b'}>385~\text{GeV}$, while the yellow one to
      $45~\text{GeV}<m_{b'}<385~\text{GeV}$. Contours of constant $m_{b'}$({\rm GeV})
      are shown in red
      dashed lines and contours of fixed $\lambda_d'\hat Y_b$({\rm GeV}) in black
      dash-dot lines. The black circle and cross show the choice of
      parameters in the examples of Sec.~\ref{sec:examples}.}
    \label{fig:plot-down-bounds}\
\end{figure}

In Fig.~\ref{fig:plot-down-bounds} we present the allowed region of parameter space for the down sector. 
The main bounds arise from the modified $Zb\bar b$ coupling  and direct searches described below.
The green region is allowed  by all measurements at $95\%$ CL while the yellow region
is model dependent.

\subsubsection{$R_b$}
According to Eq.~\eqref{Zcouplings}  $Z$-couplings are not
universal. The heavier the quark the larger the effect, so for $Z$
decays the largest and most sensitive deviation from the SM predictions is in 
$R_b$, the branching fraction to $b$ quarks. At tree level we find
\begin{equation}
\frac{\delta \Gamma_{Zb\bar b}}{ \Gamma_{Zb\bar b}}=
-s_{d_{L3}}^2\frac{2+4s_w^2Q_d}{1+4s_w^2Q_d+8s_w^4Q_d^2}\approx-2.3\,s_{d_{L3}}^2\,,
\end{equation}
and writing $\delta R_b/R_b^{\rm SM}=(1-R_b^{\rm SM})(\delta \Gamma_{b\bar b}/
\Gamma_{b\bar b}^{\rm SM})\approx 0.78(\delta \Gamma_{b\bar b}/
\Gamma_{b\bar b}^{\rm SM})$ we have
\begin{equation}
\frac{\delta R_b}{R_b^{\rm SM}}\approx -1.8 s_{d_{L3}}^2\,,
\end{equation}
to be compared to the current bound 
$\delta R_b / R_b^{\rm SM}\in[-4,8]\cdot10^{-3}$ at 95\%~CL~\cite{Amsler:2008zzb}.

Additional contributions to $\delta R_b$ from couplings to light
quarks are negligible. The virtual $t$ and $t'$
contributions  deviate from the SM's virtual $t$ contribution
by an amount that vanishes both with $m_{t'}-m_t$ and with
$s_{u_{L3}}^2$. The resulting bound on these parameters is weaker than
bounds presented below  from $V_{tb}$ (and the direct limit on $m_{t'}$).

Fig.~\ref{fig:plot-down-bounds} shows the 95\%~CL
bound from $\delta R_b$ in the $\lambda_d/y_b$ \vs\ $M_d/m_b$ plane,
where $y_b=\sqrt2m_b/v$. 

\subsubsection{Direct bounds on $m_{b'}$} 
CDF data excludes a $b'$ with mass above 100~GeV and below
268~GeV assuming $\text{BR}(b'\to Zb)=100\%$
\cite{Affolder:1999bs,Aaltonen:2007je}. For masses above
$m_{b'}=m_t+M_W=253~\text{GeV}$ the $Wt$ channel opens up and CDF data
sets a mass limit $m_{b'}> 385~\text{GeV}$ assuming $\text{BR}(b'\to
Wt)=100\%$ \cite{cdfbp}. In our model the branching fraction
assumptions may not apply. The couplings of the $b'$ to $Wt$ and $Zb$
include a suppression factor of $s_{d_{L3}}\lesssim 0.04$ (from
$R_b$). For a light Higgs the channel $b'\to bh$ can become important. 
According to Eq.~\eqref{hcouplings} the $bb'$ couplings to
the Higgs are
$\frac1{\sqrt2}\lambda_dc_{d_{L3}}c_{d_{R3}}=s_{d_{L3}}c_{d_{L3}}m_{b'}/v$
and
$\frac1{\sqrt2}\lambda_us_{d_{L3}}s_{d_{R3}}=s_{d_{L3}}c_{d_{L3}}m_b/v$.
Hence $\text{BR}(b'\to bh)$ will be large in the
region where it is kinematically allowed, provided $m_{b'}\gtrsim M_Z$.
The LEP2 95\%CL bound on the Higgs mass, $m_h>114\text{GeV}$, is valid
in this model since the properties of a light Higgs are largely
unchanged from that of the SM. Hence $100~\text{GeV}< m_{b'}\lesssim
m_b+m_h\approx 118~\text{GeV}$ is excluded. A bound $m_{b'}>128$~GeV is
given by the PDG\cite{Amsler:2008zzb} based on D0
data\cite{Abachi:1995ms} on $WW+2$jets, used in top pair production
searches. However, the bound assumes $\text{BR}(b'\to Wq)=100\%$. The
region between the LEP bound $m_{b'}\gtrsim M_Z/2$ and $m_{b'}<100$~GeV is
not easily excluded. D0 has excluded a 4th generation sequential
charge $-1/3$ quark up to $m_{b'}=m_b+M_Z$ by searching for radiative
decays $b'\to b\gamma$\cite{Abachi:1996fs} (see also
\cite{Mukhopadhyaya:1992if} for bounds using $b'\to b\ell^+\ell^-$
from analysis of Tevatron data). These bounds again may not apply in
our model, since the branching fractions are not those of a sequential
fourth generation quark. For example, the tree level three body decay
$b'\to h^*b\to b\bar b b $ can compete well with the two body
radiative decay.  The yellow and green shaded regions in Fig.~1 are
allowed by $R_b$, the thick green line labeled $Z\to b \bar b$
corresponding to the 95\% CL limit.  The green one corresponds to
$m_{b'}>385~\text{GeV}$, while the yellow one to
$45~\text{GeV}<m_{b'}<385~\text{GeV}$, and may or may not be excluded
depending on the value of the Higgs mass and, to lesser extent, other
model parameters, {\it e.g.,} flavon masses. For reference the figure
shows contours of constant $m_{b'}$ in red dashed lines and contours of
fixed $\lambda_d'\hat Y_b$ in black dash-dot lines. 

\subsection{Bounds from the up sector}

\begin{figure}[t!] 
    \centering
    \includegraphics[width=\textwidth]{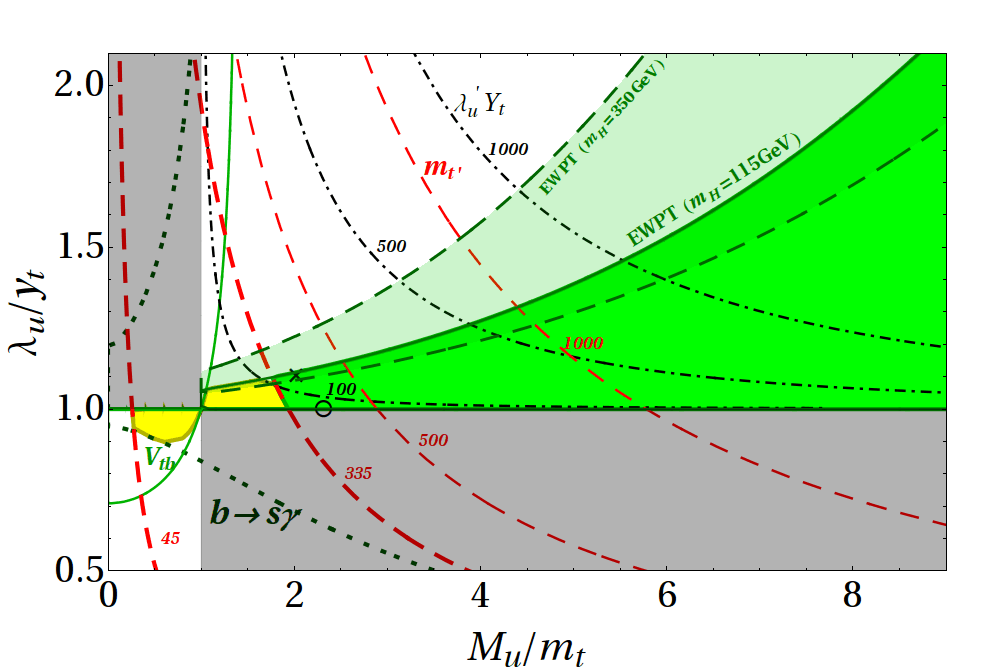}
    \caption{Allowed region of parameter space in the $\lambda_u$ \vs\  
      $M_u$ plane. The shaded grey region is unphysical. The thick green line labeled EWPT
	  shows the region allowed at 95\%~CL by the EW oblique parameters for $m_H=115~$GeV. For
	  $m_H=350~$GeV the allowed region becomes the one between the
	  green dashed lines.
	  The thin green line labeled $V_{tb}$ shows the
      95\%~CL limit from direct single top production while the green
      short-dashed line shows the 95\%~CL bound from $b\to
      s\gamma$. Of the region allowed by EWPT, $V_{tb}$ and $b\to s \gamma$
      we have distinguished $m_{t'}>335~\text{GeV}$ shaded in green from
      $45~\text{GeV}<m_{t'}<335~\text{GeV}$, shaded in yellow. For the
      latter direct mass bounds may (or not) apply, depending on the
      Higgs mass and other model parameters. Contours of constant
      $m_{t'}$({\rm GeV}) in red dashed lines and contours of fixed $\lambda_u'\hat
      Y_t$({\rm GeV}) in black dash-dot lines. The black circle and cross show the
      choice of parameters in the examples of
      Sec.~\ref{sec:examples}.}
    \label{fig:plot-up-bounds}
\end{figure}

Experimental bounds on the up sector are collected in Fig.~\ref{fig:plot-up-bounds}. The physical region of parameters
corresponds to the first and third quadrants where $m_{t'}\ge m_{t}$ and $m_{t'}\le m_{t}$, respectively. 
The main constraint in the first region arises from precision electroweak constraints and in particular from 
corrections to the $T$ parameter. The second region (where constraints from $T$, $S$ and $U$ are not applicable)
is strongly constrained by $V_{tb}$, $b\to s \gamma$ and direct searches.

\subsubsection{Electroweak Precision Tests}

The exotic fermions modify the oblique corrections to the electroweak gauge bosons 
with respect to their SM values. We compute the oblique parameters $S$, $T$ and $U$ in appendix~\ref{appendixa}.
Since the exotic fermions are $SU(2)_L$ singlets they only contribute through the mixing with SM left doublets.

After electroweak symmetry breaking the quark doublets mix with the left singlets.
This violation of custodial symmetry generates a correction to the $T$-parameter.\footnote{This was also studied recently 
in \cite{cacciapaglia} in a model with vector like top partners. For the third generation fermions 
our model reduces to theirs.} For simplicity we only consider the contributions of the
third family, which are the dominant ones. In the limit $m_b\to 0$ the exact 
one loop formula derived in App.~\ref{appendixa} reads
\begin{equation}
T=\frac{3\,s_{u_{L3}}^2}{8 \pi\, s_w^2 c_w^2} \frac{m_t^2}{M_Z^2}\left [
c_{u_{L3}}^2\,\left(\frac{m_{t'}^2}{m_{t'}^2-m_{t}^2}\log \Bigl(\frac {m_{t'}^2}{m_t^2}\Bigr)-1 \right)
+\frac{s_{u_{L3}}^2}{2} \left(\frac{m_{t'}^2}{m_t^2}-1\right) \right]\,.
\label{deltaT}
\end{equation}

As explained above for $s_{u_{L3}}=0$ the correction to $T$
vanishes. From Eq. (\ref{eq:defxyetc}) this corresponds to $M_u\to
\infty$ or $\lambda_u= y_t$ ({\it i.e.}, $\hat{Y}_u^3=0$). In the
first case the exotic fermions acquire an infinite vector like mass so
the correction to $T$ obviously vanishes in this limit.  In the second
case the mass of the top partner can be light. Since the amount of
custodial symmetry breaking is proportional to $\lambda_u$ we expect
$T$ to have the same sign as  $\lambda_u-y_t$, as is readily
checked using the explicit formulas. Even though the contribution to
$T$ is smaller than in a fourth generation model, it can be sizable and 
gives one of the most important bound on the parameters of the model. 

On the other hand, the contribution of the exotic fermions to the
$S$-parameter is always small and its sign is not fixed. 
This is similar to four-generation models, where
corrections to $S$ are generically smaller than to $T$. 
Despite the small contribution to $S$ the bound obtained combining $S$ and $T$ is significantly more restrictive that the one from $T$ alone,
due to the correlation between $S$ and $T$ in the electroweak fit. In the allowed region in Fig.~\ref{fig:plot-up-bounds} 
we have also included the bound from the $U$ parameter which however only affects the results in a minor way. 

A few words of caution. The new physics contributions to precision
electroweak parameters are here obtained as the difference between
our model and the SM one loop value. The mixing also modifies the two
loop SM correction which is not negligible in the SM. This effect is
relevant in the region where $s_{u_{L3}}$ is large which is
however only allowed in the region of small $m_{t'}$
($\ll~$TeV). In this region, however, the canonical
$S,T,U$ parameters are in general insufficient and a more refined
analysis is needed. Moreover our bounds are obtained with the assumption of a light Higgs. 
In the SM, increasing the Higgs mass would worsen the global electroweak fit mainly because of
the negative contribution to the $T$ parameter (the contribution to $S$ is instead smaller and positive, and thus well within the bound).
Interestingly in our model the correction to $S$ is always small while the contribution to $T$ is positive 
and easily of the right order to accommodate also an heavy Higgs. In Fig. 2 we have also shown the region
of parameter space allowed for a Higgs with $m_H=350~$GeV which requires a non zero mixing.
Therefore the bounds from oblique parameters should be taken with a grain of salt since
both the Higgs mass and other new physics not related to flavor may alter the bounds.

\subsubsection{$V_{tb}$}
The effective CKM matrix in Eq.~\eqref{CKMeff} is not unitary. Unitarity of the CKM matrix 
is presently only tested with significant accuracy on the first row,
$\sum_{q=d,s,b}|V^{\text{CKM}}_{uq}|^2=0.9999\pm0.0011$
\cite{Amsler:2008zzb}. However, since only light quarks participate in
this,  the resulting bound is very weak, $M_{u,d}$
greater than a few GeV.

Unitarity of the third row is more restrictive. The
measured smallness of $|V^{CKM}_{td}|$ and $|V^{CKM}_{ts}|$ implies
that the unitary matrix $V$ in Eq.~\eqref{CKMeff} has $|V_{tb}|=1$ to
high accuracy. Hence direct measurement of the $tbW$ coupling
constrains $c_{u_{L3}}$. Single top production experiments at the
Tevatron set a 95\%CL bound $|V^{\text{CKM}}_{tb}|\approx
c_{u_{L3}}>0.77$ \cite{Group:2009qk}. The resulting constraint on the
model parameters is shown in Fig.~\ref{fig:plot-up-bounds}. 
The allowed values for $c_{u_{L3}}$ at 95\% CL 
lie between the green line labeled $V_{tb}$ and the one at $\lambda_u/y_t=1$.

\subsubsection{$b\to s \gamma$}
There are two distinct underlying processes that give rise to
radiative $B$ decays. On the one hand there are $\Delta B=-\Delta S=\pm1$
operators, such as $(\bar sb)(\bar b b)$, produced by the exchange of
either a flavor vector meson or a radial mode associated to the
flavons. These contributions are highly model dependent and also very
suppressed by the overall coefficient of the four quark
operator.

On the other hand there are sizable and less model dependent
contributions from SM-like graphs with a virtual $Wt$ or $Wt'$. For
$m_{t'}>m_t$, their
sum is always larger than that of the SM. To see this note that if the
SM amplitude is a function $f(m_t)$ then the corresponding
amplitude in this model is $c^2_{u_{L3}}f(m_t)+ s^2_{u_{L3}}f(m_{t'})=
f(m_t) + s^2_{u_{L3}}(f(m_{t'})-f(m_t))$. Since $f(m_t)$ is a
monotonically increasing function, the deviation from the SM result,
$s^2_{u_{L3}}(f(m_{t'})-f(m_t))$ has the same sign as $m_{t'}-m_t$. 

In more detail, working at NLO in the simplified but accurate
approximation of Ref.~\cite{Grinstein:1990tj} we find for the process
$b\to s\gamma$ that 
\begin{equation}
\frac{\delta\Gamma_{bs\gamma}}{\Gamma_{bs\gamma}}= 2 s_{u_{L3}}^2
\frac{A(x')-A(x)-\frac83(1-z^\frac2{23})(D(x')-D(x))}{A(x)
-\frac83(1-z^\frac2{23})D(x)-\frac6{19}X_2(1-z^\frac{19}{23})},
\end{equation}
where $x=m_t^2/M_W^2$ and $x'=m_{t'}^2/M_W^2$ are arguments of the
loop functions $A$ and $D$ (given in Ref.~\cite{Grinstein:1990tj})
$z=\alpha_s(m_b)/\alpha_s(M_W)$ and $X_2=232/81$ is the coefficient of
anomalous dimension mixing the four quark operator into the transition
magnetic moment operator. The resulting 95\%~CL bound in the
$\lambda_u/y_t$ \vs\ $M_u/m_t$ plane, where $y_t=\sqrt{2}m_t/v$ is shown
as a green short-dashes line  in Fig.~\ref{fig:plot-up-bounds}. 

\subsubsection{Bounds from $m_{t'}$}
Fig.~\ref{fig:plot-up-bounds} also shows, as red dashes, contours of
fixed $m_{t'}$. CDF excludes $m_{t'}<335~\text{GeV}$ at 95\%CL, assuming
$\text{BR}(Wq)=100\%$ \cite{cdftp}. As discussed above for  the case of the $b'$
the branching fraction assumptions in the experimental analysis may not
apply in this model.  Of the region
allowed by EWPT, $V_{tb}$ and $b\to s \gamma$ we have therefore
distinguished $m_{t'}>335~\text{GeV}$ shaded in green from 
$45~\text{GeV}<m_{t'}<335~\text{GeV}$, shaded in yellow. For the latter
direct mass bounds may (or not) apply, depending on the Higgs mass and
other model parameters. For
reference the figure shows contours of constant $m_{t'}$ in red dashed
lines and contours of fixed $\lambda_u'\hat Y_t$ in black dash-dot
lines.

\subsection{Neutron EDM}
The interactions among quarks due to flavor-vector or flavon exchange
can give contributions to the Electric Dipole Moment (EDM) of
hadrons. In the SM the dominant mechanism for EDM of the neutron is
from a $\Delta S=-1$ CP violating transition $n\to \Lambda,\Sigma^0$
followed by a $\Delta S=1$ transition $\Lambda,\Sigma^0\to n \gamma$
\cite{Gavela:1981sk}. The CP violating interaction is a 1-loop induced
four-quark ``penguin'' operator
\begin{equation}
\mathcal{L}_{\text{CPV}}=i\frac{3\alpha_sG_F}{9\sqrt2\pi}
    \ln(m_t^2/m_c^2)  \text{Im}\,(V_{td}^*V_{ts}) (\bar d_L\gamma^\mu
      T^as_L)\sum_{q=u,d,s}(\bar q \gamma_\mu T^a q).
\end{equation}
A recent estimate  gives\cite{Pospelov:2005pr} 
\begin{equation}
d_n^{\text{SM}}\simeq 10^{-32} e~\text{cm}.
\end{equation}
A rough estimate of the new contributions to the neutron EDM induced by
flavor-vector or flavon exchange is obtained by replacing the
coefficient of the four-quark ``penguin'' operator in the calculation
of $d_n^{\text{SM}}$ by the CP violating coefficient $C_{\text{CPV}} $ of a
newly induced four-quark operator. One then has
\begin{equation}
\Delta d_n \sim \frac{C_{\text{CPV}} }{\frac{3\alpha_sG_F}{9\sqrt2\pi}
      \ln(m_t^2/m_c^2) \text{Im}\,(V_{td}^*V_{ts})}d_n^{\text{SM}}\approx
    7.6\times10^9 \text{GeV}^{2}\, C_{\text{CPV}} \, d_n^{\text{SM}}\,.
\end{equation}
The resulting bounds on the model parameters are extremely weak. For
example, the CP violating part of coefficients of $\Delta S=\pm1$ four-quark
operators in Eq.~(\ref{eq:4fermop}) are numerically of
order $C_{\text{CPV}} \sim10^{-15}\text{GeV}^{-2}$. The smallness of this result
justifies the crude nature of the estimate (in which we have ignored,
for example, the different possible Dirac and color structures that
may arise in the four-quark operators). The coefficients of operators
from flavon exchange, although more model dependent, are similarly
small. 

Additional contributions arise from graphs involving only electroweak
interactions but in which the heavy quarks participate. These are all
at best of the order of the SM contributions (for example, by
modifying the coefficient of the penguin operator). 

We conclude then that this class of models predicts small EDMs of
hadrons, comparable in order of magnitude to those of the SM.

\section{Signatures}
Despite the small number of extra parameters beyond the SM ones
and the relatively rigid structure of the spectrum of the model,
mostly fixed by the SM Yukawa couplings, the phenomenology above
the production threshold of new states is very rich, drastically changing
in different regions of the parameters space. We will not attempt to cover here this subject, 
which deserves a separate study. Instead we will only give a sampling of some possible 
new signatures of the model (for recent more detailed analysis on similar models
see, \eg,~\cite{AguilarSaavedra:2009es,cacciapaglia,tait},
keeping in mind however that the BR in our
case could be altered by the presence of the extra vector and flavon fields). 

Among all new states, the one that presents less model dependence is
the $b'$. As discussed before, existing searches do not provide 
very strong bounds on such particles and, as shown explicitly in the 
next section, it is easy to find parameters of the model where such resonance 
is within the reach of hadron colliders. 
The strong bounds from $Z\to b\bar b$ force the $b'$ to have small mixing
with the SM $b$ quark, which implies a small coupling with the $SU(2)_L$ gauge fields.
It turns out that, choosing $O(1)$ values for the couplings of the model (such
as $\lambda_{d}^{\phantom{\prime}}$, $\lambda'_d$ and the gauge couplings), the standard 
fourth-generation channels $Wt$, $Z b$ and $W t'$ can compete with others such 
as $Z'b$, $\tilde b b$ and $bh$. In particular the BR to $bh$ can easily be of order one.
Being a colored object, the $b'$ could be  pair produced copiously  at the LHC, provided its mass is not too high.
The signature would be quite striking having up to six bottom quarks in the final state 
($p\bar p\to \bar b' b'+X\to 2h+2b+X\to 6b+X$).

For the $t'$ the discussion is similar, with the possibility however of a substantial difference.
In this case the bound on the mixing angle $s_{u_{L3}}$ is weaker, coming only from EWPT.
Choosing as before $O(1)$ values for the parameters, we have two means
of maintaining $s_{u_{L3}}$ below the bounds,
either by increasing $M_u$ with respect to $m_t$ or by suppressing ${\hat Y}_u^3$.
In the first case we get a similar result to that of the $b'$, with the $t'\to th\to W+3b$
decay channel becoming important. The $t'$ could be pair produced at hadron
colliders, leading to very clean $6b+WW$ signals (see \cite{AguilarSaavedra:2009es} for a detailed study). 
Notice that both in the $b'$ and in the $t'$ case the $O(1)$ BR
into $hb$ and $ht$ could substantially increase the Higgs production cross section, improving the capability
of discovering and studying its properties.
In the second case, in the limit of small ${\hat Y}_u^3$, also the right mixing angle $c_{u_{R3}}$ get suppressed.
In this case the dominant channel becomes $t\tilde t$, if the radial mode of the top Yukawa ($\tilde t$) is light enough.
Actually in the limit ${\hat Y}_u^3\to 0$ the $t'$ almost decouples from the SM, 
and an approximate discrete symmetry (similar to R-parity) prevents the lightest among the $t'$ and the $\tilde t$ from decaying
into SM particles.\footnote{Actually a similar limit is also possible for the down sector, when ${\hat Y}^3_d\to 0$, $s_{d_{L3}}\to0$
and the $b'$ decouples from the $b$; however this happen in the small coupling limit 
$\lambda_d\to y_b\approx 1/40$, then the scales of the $s'$ and $d'$ decrease
accordingly and FCNC induced by the first generation may start
becoming important.} 
Such discrete symmetry is broken only by the mixing of the $t'$ sector with the other generations, 
thus producing a highly suppressed decay rate for the $t'$ in this scenario.

Finally some comments on the possible lightest gauge flavor fields.
The lightest state is expected to couple more to the third generation, and in particular to the top.
The actual couplings, however, depend on which flavor group is gauged
and the magnitude of its coupling constants.

For the lightest states three possibilities are favored. If the gauge group is $SU(3)^3$ then
the lightest state couples through the diagonal Gell-Mann $\lambda^8$ generator of $SU(3)^3$, thus with
doubled strength to the third generation respect to the first two. In this case we get 
a leptophobic non-universal $Z'$, which can be produced directly via $q\bar q$ annihilation and can decay either
into $t\bar t$ or into two jets. Existing Tevatron studies of similar
$Z'$ set mass bounds below a TeV \cite{Tevatronttbar}.
More possibilities arise when  the $U(1)$s  are also gauged. In particular the flavor boson
can mix with the SM hypercharge vector and acquire a coupling to leptons too. If such mixing
is large, the lightest vector behave as a heavy $Z'$ coupled to the hypercharge current and with
an anomalous coupling to the right-handed top. In this case strong bound are present from the EWPT \cite{Salvioni:2009mt}.
If instead the mixing with the hypercharge is negligible, then the lightest gauge boson 
will only couple to $t'_L$ and to a linear combination of $t_R$ and $t'_R$ (depending on $s_{u_{L3}}$).
In this case the four-top(top$'$) signal becomes one of the most interesting 
(see, \eg,~\cite{Brooijmans:2010tn}).

\section{Examples}
\label{sec:examples}
The details of a particular realization of the mechanism described in
Sec.~\ref{sec:InvHi} depend strongly on the actual model and
parameters chosen. Depending on the gauge group ($U(3)^3$,
$SU(3)^3\times U(1)^2$, $SU(3)^3$, $SU(3)$, $U(1)^n$,\dots), the
number and representations of scalar flavon fields and the different
parameters of the Lagrangian, the spectrum of the new particles and
their couplings may vary substantially.  Still there are some features
that are rather model independent and characterize the model.

As shown before, with the exception of the top quark sector, the
structure of the fermionic part of the model is quite rigid, depending
only on the two scales $M_u$ and $M_d$, the rest being fixed by the SM
Yukawa couplings.
Once the gauge group and the scalar content has been chosen so is the
basic structure of the spin-1 sector. But as a result of the larger
number of parameters connecting its spectrum and couplings to the SM
Yukawa terms, such as the gauge couplings and extra Yukawa couplings
($\lambda_{u,d}$, $\lambda'_{u,d}$), it is far from being specified in detail.

In the following we will provide two explicit examples where
all the parameters have been fixed, in order to demonstrate how easy
it is to build explicit models with $O(1)$ couplings, new flavor
non-universal states at the TeV scale and compatibility with all
existing experimental bounds.
In fact, depending on the choice of the parameters the strongest
bounds may come from different sources, such as EWPT, $Z\to b\bar b$, single
top production at Tevatron, $Z'$ searches and other direct bounds for
spin-1 and spin-1/2 particles, $\Delta M_K$, etc...

The two examples below correspond to the two different flavor
gaugings $SU(3)^3$ and $SU(3)^3\times U(1)^2$, respectively. For
definiteness in both cases the flavon content have been chosen to be
minimal: just the two $Y_u$ and $Y_d$ fields of
Sec.~\ref{sec:InvHi}. The couplings have been chosen to be $O(1)$ and
the two mass scales $M_u$ and $M_d$ to be low enough to produce
interesting physics for high-energy colliders and possibly for next
generation flavor experiments.

\subsection{First example: An $SU(3)^3$ model}
\begin{figure}[t!]
\begin{center}
\includegraphics[width=\textwidth]{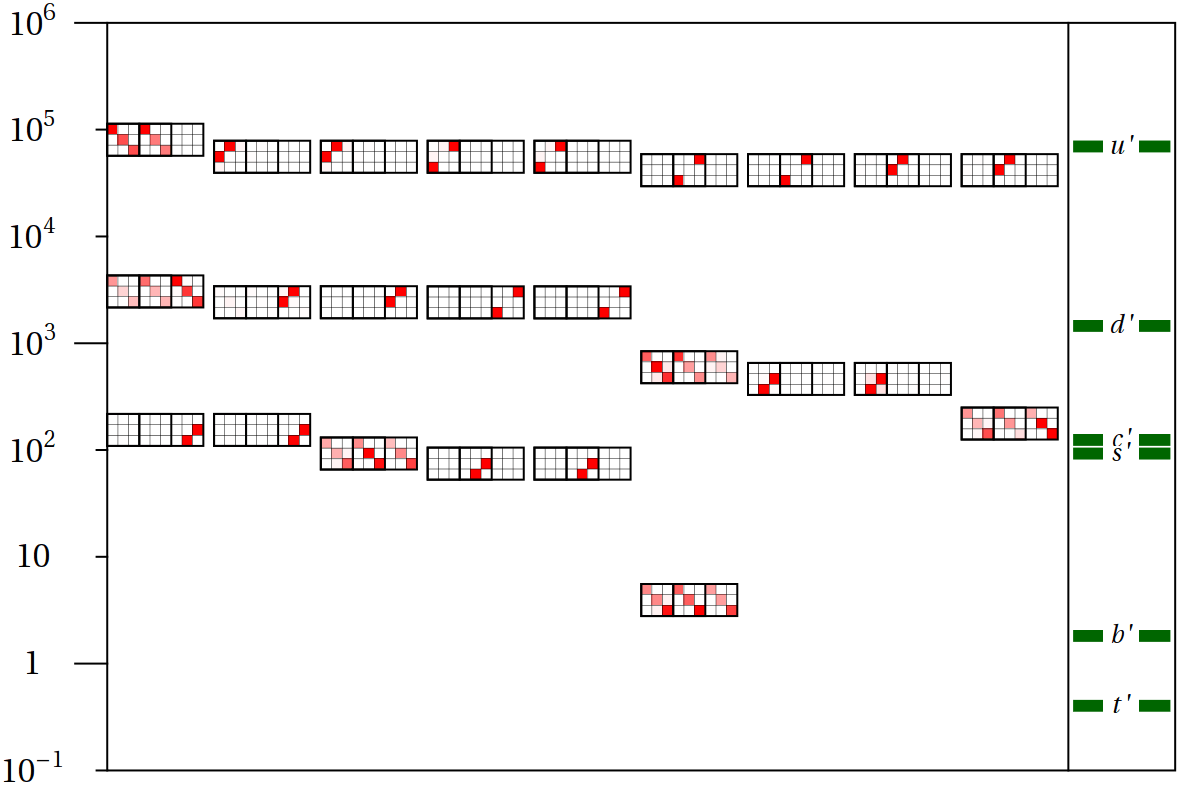}
\end{center}
\caption{\it \small \label{fig:spectrum1} Spectrum of the flavor spin-1 (left) and spin-1/2 (right) fields for the first
example (see text for details). Each vector fields is represented by a set of three $3\times3$ matrices representing
the associated generators to the three gauged $SU(3)$ groups ($SU(3)_Q$, $SU(3)_U$, $SU(3)_D$ respectively), the intensity
of the color (from white to red) correspond to the size of each entry in the generators (from 0 to 1). The position in the vertical
axis represent instead the corresponding mass in {\rm TeV}, analogously for the masses of the heavy quark partners, on the right. }
\end{figure}
In the first example we choose the following parameters:
\begin{center}
\begin{tabular}{|c|c|c|c|c|c|c|c|c|}
\hline 
$M_u$~(GeV) & $M_d$~(GeV) & $\lambda_u$ & $\lambda'_u$ & $\lambda_d$ & $\lambda'_d$ & $g_Q$ & $g_U$ & $g_D$ \\ \hline
$400$ & $100$ & $1$ & $0.5$ & $0.25$ & $0.3$ & $0.4$ & $0.3$ & $0.5$ \\  \hline
\end{tabular}
\end{center}
Given the parameters above the entries of the flavon VEVs are
fixed by requiring the right SM Yukawa couplings be reproduced, this
gives\footnote{The values of the $Y_{u,d}$ VEVs (and the the results that follow) have been calculated taking into account the running of the Yukawa couplings only up to the TeV scale. The effects coming from the running from the TeV scale up to the flavor breaking scales are more model dependent and affect mainly the value of the highest $Y_{u,d}$ VEVs, which we do not need to know with high accuracy. In fact the knowledge of the order of magnitude for these quantities  is enough for our purposes.}:
\begin{equation}
\begin{aligned}
Y_u&\approx{\rm Diag}\left (1\cdot 10^5\,,\ 2\cdot 10^2\,,\ 8\cdot 10^{-2} \right )\cdot V~{\rm TeV}\,, \\
Y_d&\approx{\rm Diag}\left (5\cdot 10^3\,,\ 3\cdot 10^2\,,\ 6 \right )~{\rm TeV}\,,
\end{aligned}
\end{equation}
where $V$ is the unitary experimental CKM matrix \cite{Amsler:2008zzb}.

The couplings are chosen to be smaller than 1 to avoid possible problems
with early Landau-poles except for $\lambda_u$, which must be larger
than $y_t=\sqrt2 m_t/v\simeq1$ (or slightly smaller when $m_{t'}<m_t$; see Sec.~\ref{sec:modsmcoup}). 
For $\lambda_u=1$, as in this example, the mixing of the
left doublet is small and the lowest eigenvalue of $Y_u$ approaches zero.

Given the parameters above we can calculate both the spectrum and
couplings of the spin-1 and spin-1/2 sectors of the theory. The
spectrum is summarized in Fig.~\ref{fig:spectrum1}.

The masses of the four lightest spin-1 states are $2.8$, $53$, $53$,
and $66$~TeV.  The lightest state, which is one order of magnitude
lighter than the next to lightest one, couples to fermions through the
$\lambda^8$ flavor generator and with equal strength to left/right
up/down type fermions (the unequal intensity of shading in Fig.~\ref{fig:spectrum1} is  compensated by the different values of
the gauge couplings).  Although its coupling to
the third generation is the largest, the lightest vector couples also
to the first two generations, which makes it accessible at the LHC. For
all practical purposes it corresponds to a flavor non-universal
leptophobic $Z'$. The existing mass bounds on analogous resonances 
from Tevatron searches in the $t\bar t$ channel lie below 1~TeV \cite{Tevatronttbar}.

The masses of the three lightest fermion fields are  $0.40$,
$1.8$, and $90$~TeV.  In this case both lightest states, the
$t'$ and the $b'$, should be within the reach of the LHC. It is
important to keep in mind however that, contrary to 4th generation quark
fields, their couplings to the SM $W$ and $Z$ bosons arise through  
mixing with SM left-handed fields and are suppressed by
the small angles $s_{u_{L3}}$ and $s_{d_{L3}}$, which   
for the current choice of parameters are $0.05$ and $0.02$, respectively.

It is interesting to check how much this model is actually safe
against existing bounds. The values of parameters chosen for this
case correspond to the point with symbol ``{\Large $\circ$}'' in the
($\lambda_{u/d}$, $M_{u/d}$) planes of Figs.~\ref{fig:plot-down-bounds}
and \ref{fig:plot-up-bounds}. The contributions to $Z\to b\bar b$, EW precision observables and $V_{tb}$ read
\begin{equation}
\begin{aligned}
\frac{\delta R_{b}}{R_b}&=-1.0 \cdot 10^{-3}\,, \\ 
S&=0.00\,,\ T=0.01\,,\ U=0.00\,, \\
 V_{tb}&=1.00\,.
\end{aligned}
\end{equation}
Except for the correction to $R_b$ which is naturally suppressed by the $b$ mass
the small corrections to the observables above are due to the choice 
$\lambda_u\simeq y_t$. In this region of parameters, which arises automatically anytime 
$\lambda' _u{\hat Y}_u^{3}\ll M_u$,  the new physics in the up-sector
decouples from the SM. 

As discussed in section~\ref{sec:VandS} the exchange of flavor gauge
bosons can also produce  flavor breaking 4-fermion operators at tree
level. Existing strong bounds on these operators are often used to
rule out the possibility of  low scale flavor vector fields.
However the inverted hierarchy present in our spectrum allows to
easily avoid all such bounds.  Indeed,  from
Fig.~\ref{fig:spectrum1} we note that the vector fields mediating transitions
among the first and the higher generations are among the heaviest,
followed by those mediating transition among the second and the third
generations, while the lightest is flavor diagonal.  At tree level the
strongest bounds come from $\Delta F=2$ quark transitions, whose
bounds on 4-fermion operators are conveniently summarized in
\cite{Bona:2007vi}. Our vector boson only produce three types of such
operators at tree level:
\begin{equation}
\begin{aligned}
Q_1^{q_i q_j}&={\overline q}^\alpha_{jL} \gamma_\mu q_{iL}^\alpha\,{\overline q}^\beta_{jL} \gamma^\mu q_{iL}^\beta \,,\\
\tilde Q_1^{q_i q_j}&={\overline q}^\alpha_{jR} \gamma_\mu q_{iR}^\alpha\,{\overline q}^\beta_{jR} \gamma^\mu q_{iR}^\beta \,,\\
Q_5^{q_i q_j}&={\overline q}^\alpha_{jR} q_{iL}^\beta\,{\overline q}^\beta_{jL} q_{iR}^\alpha \,.
\end{aligned}
\end{equation}
The coefficients of these operators can be obtained numerically from eq. (\ref{eq:4fermop}). In our explicit example 
they read:
\begin{center}
\begin{tabular}{||c||c|c||}
\hline 
& Re (in GeV$^{-2}$) & Im (in GeV$^{-2}$) \\ \hline
$C^1_K$ 			& $-1\cdot 10^{-14}$ & $-1\cdot 10^{-19}$ \\
$\tilde C^1_K$ 		& $-2\cdot 10^{-16}$ & $-2\cdot 10^{-21}$\\
$C^5_K $			& $-5\cdot 10^{-15}$ & $-6\cdot 10^{-20}$ \\ \hline
$C^1_D $			& $-2\cdot 10^{-20}$ & $-2\cdot 10^{-23}$\\
$\tilde C^1_D$ 		& $-2\cdot 10^{-25}$ & $-2\cdot 10^{-28}$\\
$C^5_D $			& $-2\cdot 10^{-22}$ & $-2\cdot 10^{-25}$\\ \hline
$C^1_{B_d}$       	& $ 1\cdot 10^{-16}$ & $5\cdot 10^{-16}$\\
$\tilde C^1_{B_d}$	& $ 9\cdot 10^{-22}$ & $3\cdot 10^{-21}$\\
$C^5_{B_d}$ 		& $ 1\cdot 10^{-18}$ & $5\cdot 10^{-18}$\\ \hline
$C^1_{B_s}$ 		& $ 3\cdot 10^{-13}$ & $-4\cdot 10^{-13}$ \\
$\tilde C^1_{B_s}$	& $ 4\cdot 10^{-16}$ & $-6\cdot 10^{-16}$ \\
$C^5_{B_s}$ 		& $ 4\cdot 10^{-14}$ & $-6\cdot 10^{-14}$ \\ \hline
\end{tabular}
\end{center}
We have used here the notation for coefficients of
Ref.~\cite{Bona:2007vi}. Comparing with the bounds in that work,
\begin{center}
\begin{tabular}{||c||c|c||}
\hline 
& Re (in GeV$^{-2}$) & Im (in GeV$^{-2}$) \\ \hline
$C^1_K$ 			& $[-9.6,9.6]\cdot 10^{-13}$ & $[-4.4,2.8]\cdot 10^{-15}$ \\
$\tilde C^1_K$ 		& $[-9.6,9.6]\cdot 10^{-13}$ & $[-4.4,2.8]\cdot 10^{-15}$ \\
$C^5_K $			& $[-1.0,1.0]\cdot 10^{-14}$ & $[-5.2,2.9]\cdot 10^{-17}$ \\ \hline
$|C^1_D|$			& \multicolumn{2}{c||}{$<7.2\cdot 10^{-14}$} \\
$|\tilde C^1_D|$ 	& \multicolumn{2}{c||}{$<7.2\cdot 10^{-14}$} \\
$|C^5_D| $			& \multicolumn{2}{c||}{$<4.8\cdot 10^{-13}$} \\ \hline
$|C^1_{B_d}|$      	& \multicolumn{2}{c||}{$<2.3\cdot 10^{-11}$} \\
$|\tilde C^1_{B_d}|$& \multicolumn{2}{c||}{$<2.3\cdot 10^{-11}$} \\
$|C^5_{B_d}|$ 		& \multicolumn{2}{c||}{$<6.0\cdot 10^{-13}$} \\ \hline
$|C^1_{B_s}|$ 		& \multicolumn{2}{c||}{$<1.1\cdot 10^{-9}$}  \\
$|\tilde C^1_{B_s}|$& \multicolumn{2}{c||}{$<1.1\cdot 10^{-9}$}  \\
$|C^5_{B_s}|$ 		& \multicolumn{2}{c||}{$<4.5\cdot 10^{-11}$} \\ \hline
\end{tabular}
\end{center}
one realizes that the resulting FCNC processes
are well within the experimental bounds, with the most dangerous one
(Re\,$C^5_K$) still a factor of two smaller than current limits.
This is so even if we chose extreme parameters that make flavorful
new physics lie just beyond the exclusion bounds from direct searches and 
from flavor non-violating observables.

\subsection{Second example: An $SU(3)^3\times U(1)^2$ model}
\begin{figure}[t!]
\begin{center}
\includegraphics[width=\textwidth]{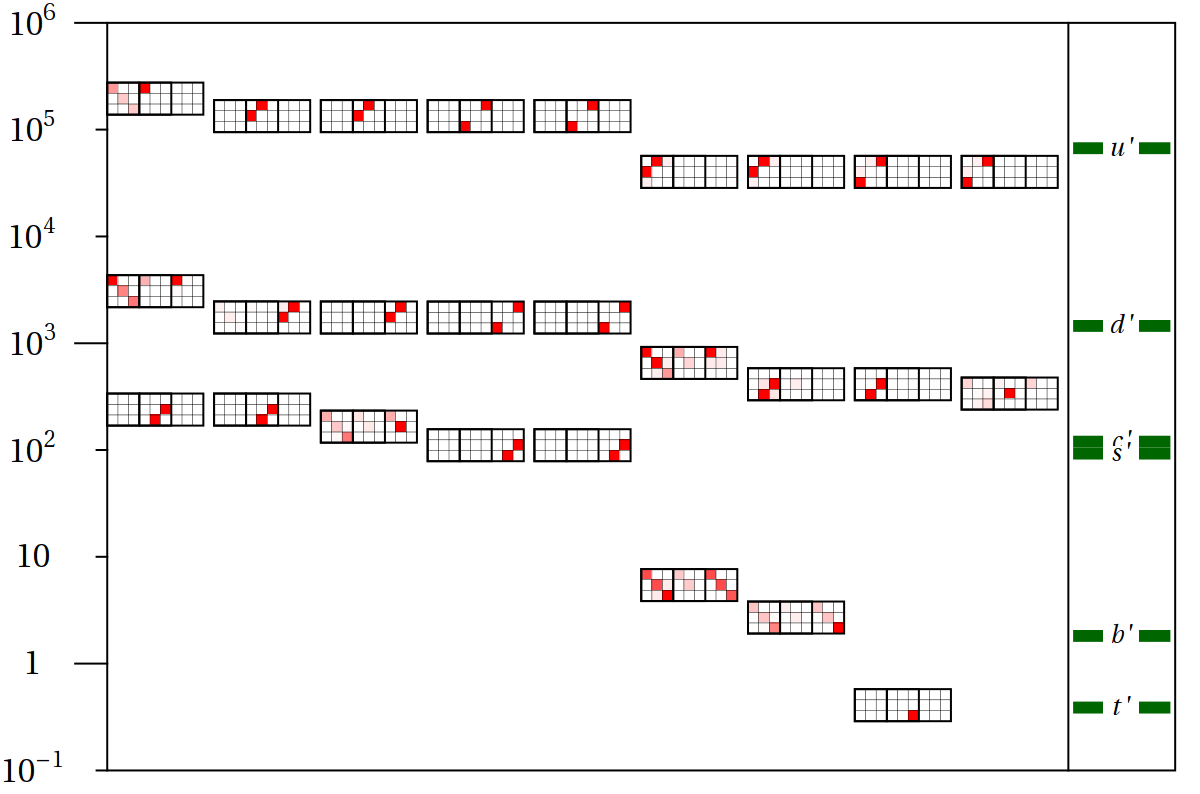}
\end{center}
\caption{\it \small \label{fig:spectrum2} Spectrum of the flavor spin-1 (left) and spin-1/2 (right) fields for the second
example (see text and caption of Fig.~\ref{fig:spectrum1} for details). }
\end{figure}
Our second example involves the gauging of $SU(3)^3\times
U(1)^2$. With respect to the previous one, two extra vector fields
have been added, corresponding to the right-handed up and down flavor
numbers.  The flavor gauge fields can thus be identified with the
generators of the $SU(3)_{Q_L}\times U(3)_{U_R}\times U(3)_{D_R}$ group. 
This means that now they are free to mix (even above the flavor breaking scale) with the SM
hypercharge. This effect produces a mixing between the flavor gauge
bosons (only those with a non-vanishing U(1) component) and the SM $Z$
boson. We thus have two extra free parameters, characterizing the
hypercharge mixing with each of the two flavor U(1). There are two
expected sizes for such mixing: a) ${\cal O}(1)$ if they started
${\cal O}(1)$ at some high scale; b) ``one-loop
suppressed''$\times$``$logs$'' if they were suppressed for some reason at
the high scale and are produced only via radiative corrections. Since the mixing
with the hypercharge does not change the flavor-breaking structure,
it will not alter significantly the calculation of the flavor
breaking effects. However the mixing with the hypercharge now allows
such vector fields to couple to leptons at tree level. On the one hand
the mixing makes it easier to detect such vector fields through their
leptonic channels. On the other, however, it makes the bound on their
masses stronger, in order to escape limits from electroweak precision
tests.

The values of parameters we choose in this example are similar to
those of the previous one:
\begin{center}
\begin{tabular}{|c|c|c|c|c|c|c|c|c|}
\hline 
$M_u$~(GeV) & $M_d$~(GeV) & $\lambda_u$ & $\lambda'_u$ & $\lambda_d$ & $\lambda'_d$ & $g_Q$ & $g_U$ & $g_D$ \\ \hline
$350$ & $100$ & $1.1$ & $0.5$ & $0.25$ & $0.25$ & $0.3$ & $1$ & $0.3$ \\  \hline
\end{tabular}
\end{center}
As before together with the values of the SM Yukawa couplings these parameters fix the values of the flavon VEVs:
\begin{equation}
\begin{aligned}
Y_u&\approx {\rm Diag}\left ( 1\cdot 10^5\,,\ 2\cdot 10^2\,,\ 3\cdot 10^{-1} \right )\cdot V~{\rm TeV}\,, \\
Y_d&\approx {\rm Diag}\left (6\cdot 10^3\,,\ 4\cdot 10^2\,,\ 7 \right )~{\rm TeV}\,.
\end{aligned}
\end{equation}

The spectrum is summarized in Fig.~\ref{fig:spectrum2}. 
As expected the fermionic spectrum is not very different from the previous example, with
the three lightest states having masses of 0.4, 1.8, and 90 TeV.

The gauge boson spectrum underwent larger
modifications. Now we have two extra states, which
populate the lowest part of the spectrum.  The four lightest states
have now the following masses: 0.29, 1.9, 3.9, and 80 TeV.  Thus we
have three flavor gauge bosons that in principle are within the reach
of the LHC.  In particular the possibility of having vector fields
associated to non-traceless generators allowed the presence of a very
light vector particle coupled only to the right-handed third 
generation charge $+2/3$ quarks, $t_R$ and $t'_R$, since it receives a mass only from the
$\hat{Y}_u^{3}$ entry of the flavon field, which is the smallest one.

Neglecting possible kinetic mixing, the lightest vector couples only  to
$t_R$. For this reason it could have escaped detection and indirect
bounds, despite its low mass. Once the mixing with hypercharge is
taken into account strong bounds from EWPT may start becoming
important: for a $Z'$ coupling to  hypercharge the bound reads:
$M_{Z'}/g_{Z'}\leq 8.55$~TeV (\@95\% CL) \cite{Salvioni:2009mt}. This
implies that in this explicit model we may allow for a mixing not
bigger than 5\% to avoid conflicts with EWPT.

The position of the parameters chosen for this example in the
($M_{u/d}$, $\lambda_{u/d}$) plane is shown with the symbol
``$\times$'' in Figs.~\ref{fig:plot-down-bounds} and
\ref{fig:plot-up-bounds}, thus within the experimental bounds coming
from direct searches of $m_{b'}$ and $m_{t'}$ and indirect effects such as
$Z\to b\bar b$, EWPT and $V_{tb}$.  
In particular, for these latter quantities we get
\begin{equation}
\begin{aligned}
\frac{\delta R_{b}}{R_b}&= - 1.0 \cdot 10^{-3}\,,  \\
S&=0.00\,,\ T=0.15\,, \ U=0.01 \,,\\
V_{tb}&=0.97\,.
\end{aligned}
\end{equation}
This is close to the bounds from EWPT, as can also be seen from Fig.~\ref{fig:plot-up-bounds}. The figure also shows that
for this values of parameters a heavy Higgs with mass up to $\sim350~$GeV  is still allowed.

Finally the effects on $\Delta F=2$ processes are (see the previous example for details):
\begin{center}
\begin{tabular}{||c||c|c||}
\hline 
& Re (in GeV$^{-2}$) & Im (in GeV$^{-2}$) \\ \hline
$C^1_K$ 			& $-7\cdot 10^{-15}$ & $-8\cdot 10^{-20}$ \\
$\tilde C^1_K$ 		& $-1\cdot 10^{-16}$ & $-1\cdot 10^{-21}$\\
$C^5_K $			& $-4\cdot 10^{-15}$ & $-4\cdot 10^{-20}$ \\ \hline
$C^1_D $			& $-3\cdot 10^{-20}$ & $-3\cdot 10^{-23}$\\
$\tilde C^1_D$ 		& $-3\cdot 10^{-25}$ & $-4\cdot 10^{-28}$\\
$C^5_D $			& $-4\cdot 10^{-22}$ & $-4\cdot 10^{-25}$\\ \hline
$C^1_{B_d}$       	& $ 2\cdot 10^{-16}$ & $2\cdot 10^{-16}$\\
$\tilde C^1_{B_d}$	& $ 1\cdot 10^{-21}$ & $1\cdot 10^{-21}$\\
$C^5_{B_d}$ 		& $ 2\cdot 10^{-18}$ & $2\cdot 10^{-18}$\\ \hline
$C^1_{B_s}$ 		& $ 3\cdot 10^{-13}$ & $-4\cdot 10^{-13}$ \\
$\tilde C^1_{B_s}$	& $ 5\cdot 10^{-16}$ & $-6\cdot 10^{-16}$ \\
$C^5_{B_s}$ 		& $ 5\cdot 10^{-14}$ & $-6\cdot 10^{-14}$ \\ \hline
\end{tabular}
\end{center}
which are similar or smaller to those of the previous example, thus well within the experimental bounds.

\section{Discussion}

We have investigated the possibility of gauging the SM flavor symmetries.
Remarkably cancellation of gauge anomalies automatically leads to a model characterized by a hierarchical 
structure of new physics where the light generations are protected from large corrections with respect to the SM
predictions, while deviations could be present for the top and bottom quarks. Contrary to the standard lore, 
the mechanism described here allows the scale of flavor physics to be as low as a TeV while
avoiding all flavor and precision electroweak bounds but within reach at the Tevatron and the LHC. 
The  lightest new states are the top partners in the fermionic sector
and a few flavor gauge bosons
that behave as non-universal $Z'$. Depending on the flavor gauge group a few flavor gauge bosons could 
be observable. Most of the spectrum is much heavier than a TeV and can
not be accessed directly in present day accelerators.
However the contributions could still be important for precision observables particularly in flavor physics. 
The actual details of the model can vary substantially (the choice of the gauge group, the number of flavon fields, 
values of coupling constants, etc.) however the general structure of inverted hierarchy is rather robust.

The main drawback of our model is that the scale of new physics, roughly set by the 
parameters $M_{u,d}$, is an arbitrary parameter which, if larger than
few TeV would render the new states heavy, out of reach 
of present experiments. We are tempted to speculate that the scale of flavor physics is linked to the 
electroweak scale implementing this mechanism within a theory that addresses the hierarchy problem in the SM.
In general flavor physics imposes formidable constraints on physics beyond the SM. At present two
strategies seem possible. The first is to demand that new physics respects a MFV structure. 
To our knowledge however this hypothesis cannot be derived from a symmetry of the UV theory 
but only arise in the IR accidentally. The other possibility is the idea of partial 
compositeness, see for example \cite{contino}. In this case the light generations are elementary as in the SM and the
flavor transitions are suppressed by the small mixing with composite states to which the Higgs couple. 
Unfortunately the flavor and CP protection achieved in this case seems at present incomplete. 
The inverted hierarchy of our model has some similarities with both scenarios but here 
the suppression is due to the large mass of the relevant degrees of freedom rather than the coupling.  
Of course some obvious challenges should be faced in particular how to avoid reintroducing 
quadratic divergences in the Higgs sector once the new physics  has hierarchical scales. 

We hope to return to this question in the future.

\vspace{0.5cm} {\bf Acknowledgments:} 
We are grateful to Gia Dvali for inspiring discussions. 
We also thank G\'eraldine Servant for discussions,
Gino Isidori and Graham Ross for comments on the manuscript and
Diego Guadagnoli for pointing out some typos in appendix~\ref{appendixa}.
The work of BG was supported in part by the US Department 
of Energy under contract DE-FG03-97ER40546.

\appendix

\section{Radial modes}
\label{appendixscalar}

Radial and GB modes of the flavon fields can be parametrized as (see also \cite{feldmann} for related discussion),
\begin{equation}
\begin{aligned}
Y_u&=U_{U} \rho_u U_{Q}^\dagger\,, \\
Y_d&=U_{D} \rho_d U_{Q}^\dagger\,,
\end{aligned}
\end{equation}
where $U_{Q,U,D}$ are the three unitary matrices parametrizing the 9+9+8=26 Goldstone modes. $\rho_u$ and $\rho_d$
are the matrices of the remaining 36-26=10 radial modes, with VEVs
\begin{equation}
\begin{aligned}
\langle \rho_u \rangle &=\frac{\lambda_u M_u}{\lambda'_u} \hat y_u^{-1} V\,, \\
\langle \rho_d \rangle &=\frac{\lambda_d M_d}{\lambda'_d} \hat y_d^{-1}\,, 
\end{aligned}
\end{equation}
where, for simplicity, we assumed small Yukawa couplings 
(for the third generation the exact expression, Eq.~(\ref{eq:defxyetc}), should be used).
Requiring that the radial modes in $\rho_{u,d}$ be orthogonal to the Goldstone modes correspond in the unitary gauge
to the condition that cross product terms of the type $\partial_\mu \rho A^\mu$ vanish. This correspond to the the three sets of conditions
\begin{equation}
\begin{aligned} 
\label{eq:orthoconstr}
A_U :&\quad {\rm Im\ Tr} [\rho_u \partial \rho_u^\dagger \lambda^\alpha]=0 \,, &\alpha&=1,\ldots,9 \\
A_D :&\quad {\rm Im\ Tr} [\rho_d \partial \rho_d^\dagger \lambda^\alpha]=0 \,, &\alpha&=1,\ldots,9 \\
A_Q :&\quad {\rm Im\ Tr} [(\partial \rho_u^\dagger \rho_u+\partial \rho_d^\dagger \rho_d)  \lambda^\alpha]=0 \,, &\alpha&=1,\ldots,8\,.
\end{aligned}
\end{equation}
We can conveniently rewrite
\begin{equation}
\begin{aligned}\label{eq:defrhoud}
\rho_u&=\Sigma_{Ru}D_u \Sigma_{Lu}^\dagger V\,, \\
\rho_d&=\Sigma_{Rd}D_d \Sigma_{Ld}^\dagger \,, 
\end{aligned}
\end{equation}
where $D_{u,d}$ are real diagonal matrices parametrizing the 6 radial modes associated to the Yukawa masses
($\langle D_{u,d}\rangle=\lambda_{u,d}M_{u,d}\hat y_{u,d}^{-1}/\lambda'_{u,d}$),
and $\Sigma_{Ru,Lu,Rd,Ld}$ are four unitary matrices parametrizing (after imposing the constraints)
the remaining four angle modes. In particular we can write
\begin{eqnarray}
\Sigma_{X}=\exp{\left (i \Pi_X \right)}=\exp{\left (i \frac{\lambda^\alpha}{2}\pi^\alpha_X\right)}\,.
\end{eqnarray}
Out of the corresponding 36 fields $\pi^\alpha_X$ only 4 combinations remain since
6 cancel out in the combination~(\ref{eq:defrhoud}) 
and 26 are removed by the constraints~(\ref{eq:orthoconstr}).
The latter in terms of the $\Pi_X$ fields read
\begin{equation}
\begin{aligned} 
\label{eq:orthoconstr2}
A_U :&\quad {\rm Tr} [(2D_u \Pi_{Lu}D_u-D_u^2\Pi_{Ru}-\Pi_{Ru}D_u^2) \lambda^\alpha]=0 \,,\qquad &\alpha&=1,\ldots,9 \\
A_D :&\quad {\rm Tr} [(2D_d \Pi_{Ld}D_d-D_d^2\Pi_{Rd}-\Pi_{Rd}D_d^2) \lambda^\alpha]=0 \,,\qquad &\alpha&=1,\ldots,9 \\
A_Q :&\quad {\rm Tr} [(V^\dagger(2D_u
\Pi_{Ru}D_u-D_u^2\Pi_{Lu}-\Pi_{Lu}D_u^2)V  & & \\
	&\quad +2D_d \Pi_{Rd}D_d-D_d^2\Pi_{Ld}-\Pi_{Ld}D_d^2)\lambda^\alpha]=0 \,,\qquad &\alpha&=1,\ldots,8\,.
\end{aligned}
\end{equation}

The combination of $\pi^\alpha_X$ which cancel out can be found from the relations
\begin{equation}
\Sigma_{Ru}D_u \Sigma_{Lu}^\dagger = D_u\,, \qquad 
\Sigma_{Rd}D_d \Sigma_{Ld}^\dagger = D_d\,,
\end{equation}
which give the following constraints for the remaining modes:
\begin{equation}
\pi_{Ru}^{3,8,9}=-\pi_{Lu}^{3,8,9}\,,\quad \pi_{Rd}^{3,8,9}=-\pi_{Ld}^{3,8,9}\,.
\end{equation}

The first 9+9 conditions of (\ref{eq:orthoconstr}) give the following constraints:
\begin{equation}
\begin{aligned}
2 d_u d_c \pi_{Lu}^{1,2}&=\pi_{Ru}^{1,2} (d_u^2+d_c^2)\,, \\
2 d_u d_t \pi_{Lu}^{4,5}&=\pi_{Ru}^{4,5} (d_u^2+d_t^2)\,, \\
2 d_c d_t \pi_{Lu}^{6,7}&=\pi_{Ru}^{6,7} (d_c^2+d_t^2)\,, \\
\pi_{Ru}^{3,8,9}&=\pi_{Lu}^{3,8,9}\,, \\
2 d_d d_s \pi_{Ld}^{1,2}&=\pi_{Rd}^{1,2} (d_d^2+d_s^2)\,, \\
2 d_d d_b \pi_{Ld}^{4,5}&=\pi_{Rd}^{4,5} (d_d^2+d_b^2)\,, \\
2 d_d d_b \pi_{Ld}^{6,7}&=\pi_{Rd}^{6,7} (d_s^2+d_b^2)\,, \\
\pi_{Rd}^{3,8,9}&=\pi_{Ld}^{3,8,9}\,,
\end{aligned}
\end{equation}
where $\langle D_{u,d}\rangle=\text{Diag}(d_{u,d},d_{c,s},d_{t,b})$ and together with the previous condition imply
that $\pi_X^{3,8,9}=0$\,.

We thus ended up with 12 fields, without lack of generality $\pi_{Ru,Rd}^{1,2,4,5,6,7}$.
There are 8 further constraints from the last line in (\ref{eq:orthoconstr2}), which leave only four independent fields.
Notice that these are the only constraints that make the CKM angles appear. The expressions we obtain are quite lengthy
and we do not report them here explicitly, but we only notice that all the 12 fields are in general different from zero
and can be written as linear combination of four independent fields.

\subsection{General facts about radial modes}
From the parametrization given above we notice some interesting facts about the way the radial modes couple.
Among the invariants that can be written in the Lagrangian those which are only functions of one type
of flavon field,  depend only on the diagonal modes in a simple way. Indeed
\begin{equation}
\begin{aligned}
{\rm Tr}[(Y_{u,d}^\dagger Y_{u,d})^n]&={\rm Tr}[(D_{u,d}^{2n}]\,, \\
{\rm Det}[Y_{u,d}]&={\rm Det}[D_{u,d}]\,.
\end{aligned}
\end{equation}
$\Sigma$-flavons only appear when both $Y_u$ and $Y_d$ are present.
We have for example
\begin{eqnarray}
{\rm Tr}[Y_{u}^\dagger Y_{u}Y_{d}^\dagger Y_{d}]=
{\rm Tr}[D_{u}^2(\Sigma_{Lu}V\Sigma_{Ld}^\dagger)D_d^2(\Sigma_{Lu}V\Sigma_{Ld}^\dagger)^\dagger]\,,
\end{eqnarray}
and in general the operators will be strings of the type
\begin{equation}
{\rm Tr} [D_u^{2n_1} (\Sigma_{Lu}V\Sigma_{Ld}^\dagger)D_d^{2n_2} (\Sigma_{Lu}V\Sigma_{Ld}^\dagger)^\dagger
D_u^{2n_3} (\Sigma_{Lu}V\Sigma_{Ld}^\dagger)^\dagger \dots ]\,.
\end{equation}
Thus the CKM radial modes only enter through the combination $(\Sigma_{Lu}V\Sigma_{Ld}^\dagger)$.

In the coupling to the SM fermions we have instead
\begin{equation}
\begin{aligned}
{\overline Q} \tilde H \frac{\lambda_u M_u}{\lambda'_u} Y_{u}^{-1} U_R 
	&= {\overline Q} \tilde H \frac{\lambda_u M_u}{\lambda'_u} V^\dagger \Sigma_{Lu} D_{u}^{-1} \Sigma_{Ru}^\dagger U_R \,, \\
{\overline Q} H \frac{\lambda_d M_d}{\lambda'_d} Y_{d}^{-1} D_R 
	&= {\overline Q} H \frac{\lambda_d M_d}{\lambda'_d} \Sigma_{Ld} D_{d}^{-1} \Sigma_{Rd}^\dagger D_R\,.
\end{aligned}
\end{equation}
Calculating only three particle vertices, relevant for tree-level flavor breaking, we have two possible types of operators,
from the interaction of SM fermions to the diagonal and the CKM radial modes respectively.
For the first we can put the Higgs and the CKM modes to their VEVs $\Sigma_X=1$ and get
\begin{equation}
\begin{aligned}
\frac{v}{\sqrt2}{\overline U_L}  V^\dagger  \frac{\lambda_u M_u}{\lambda'_u}  D_{u}^{-1}  U_R &\to
\frac{v}{\sqrt2}{\overline U_L}   \frac{\lambda_u M_u}{\lambda'_u}  D_{u}^{-1}  U_R 	=-\frac{\sqrt2 \lambda'_u}{\lambda_u}\frac{m_{u^i}^2}{M_u v}{\overline U}_L^i   D_u^{ii}  U_R^i \,, \\
\frac{v}{\sqrt2}{\overline D_L}    \frac{\lambda_d M_d}{\lambda'_d}  D_{d}^{-1}  D_R 
	&=-\frac{\sqrt2 \lambda'_d}{\lambda_d}\frac{m_{d^i}^2}{M_d v}{\overline D}_L^i   D_d^{ii}  D_R^i \,,
\end{aligned}
\end{equation}
where we went from the Yukawa to the quark mass eigenstate basis $U_L \to V U_L$.
We make two observations here. First, the interactions of these modes
are doubly suppressed by the Yukawa coupling constants (one suppression
more than for the Higgs), and second, the interactions are flavor diagonal in the mass eigenstate basis, a sort of GIM mechanism
is at work and they do not induce FC processes at tree level.

The interactions of the CKM modes read instead
\begin{equation}
\begin{aligned}
{\overline U}_L^i \Sigma_{Lu}^{ij} m_{u_j} (\Sigma_{Ru}^\dagger)^{jk} U_R^k
	=i{\overline U}_L^i (\Pi_{Lu}^{ij} m_{u_j} -m_{u_i}\Pi_{Ru}^{ij}) U_R^j \,, \\
{\overline D}_L^i \Sigma_{Ld}^{ij} m_{d_j} (\Sigma_{Rd}^\dagger)^{jk} D_R
	=i{\overline D}_L^i (\Pi_{Ld}^{ij} m_{d_j} -m_{d_i}\Pi_{Rd}^{ij}) D_R^j\,,
\end{aligned}
\end{equation}
from where we see that they can mediate flavor violations. 
Alternatively, the $\Sigma$-fields can be reabsorbed into a field redefinition of the quark fields,
which makes the interactions appear from the kinetic terms:
\begin{equation}
\begin{aligned}
i{\overline U}_{L,R} \gamma^\mu \Sigma_{Lu,Ru}^\dagger \partial_\mu \Sigma_{Lu,Ru} U_{L,R} =
	-{\overline U}_{L,R} \gamma^\mu \partial_\mu \Pi_{Lu,Ru} U_{L,R} \,, \\
i{\overline D}_{L,R} \gamma^\mu \Sigma_{Ld,Rd}^\dagger \partial_\mu \Sigma_{Ld,Rd} D_{L,R} =
	-{\overline D}_{L,R} \gamma^\mu \partial_\mu \Pi_{Ld,Rd} D_{L,R} \,.
\end{aligned}
\end{equation}
In this form the interactions of the CKM modes resembles the one of the longitudinal modes of
the vector fields. To  estimate the potential flavor violation induced by these
interactions, we should write explicitly the dependence on the independent modes in the $\Pi$-fields
and work out the spectrum of the corresponding modes. The full analytic expression turns out to be
lengthy and not very illuminating, therefore we will give them explicitly in the two-flavor case
to illustrate their structure, while for the three-flavor case 
we will give only the numerical estimates.

\subsection{The 2-flavors example}
In the two dimensional case there is only one CKM mode, which means that all $\Pi$-fields can be
rewritten in terms of one field only. The diagonal entries vanish because of the constraints,
like in the 3 flavor case. It is simple to work out all the constraints and the explicit results
for the $\Pi$-fields, in terms of the  canonically
normalized field $\varphi$,  read
\begin{equation}
\begin{aligned}
\Pi_{Lu}&=\frac{\sigma^2}{2}\frac{d_d^2-d_s^2}{d_u^2-d_c^2}\frac{d_u^2+d_c^2}{\kappa} \varphi \,,  \\
\Pi_{Ru}&=\frac{\sigma^2}{2}\frac{d_d^2-d_s^2}{d_u^2-d_c^2}\frac{2d_u d_c}{\kappa} \varphi \,,  \\
\Pi_{Ld}&=\frac{\sigma^2}{2}\frac{d_u^2-d_c^2}{d_d^2-d_s^2}\frac{d_d^2+d_s^2}{\kappa} \varphi \,,  \\
\Pi_{Rd}&=\frac{\sigma^2}{2}\frac{d_u^2-d_c^2}{d_d^2-d_s^2}\frac{2 d_d d_s}{\kappa} \varphi \,,  \\
\kappa&=\sqrt{(d_d^2+d_s^2)(d_u^2-d_c^2)^2+(d_u^2+d_c^2)(d_d^2-d_s^2)^2}\,,
\end{aligned}
\end{equation}
where $\sigma^a$ are Pauli matrices.
As we explained earlier only the combination $(\Sigma_{Lu}V\Sigma_{Ld}^\dagger)$ can appear in the scalar
potential. Notice that in this case $V=\exp(i \sigma^2
\theta_{12})$ and therefore
\begin{equation}
(\Sigma_{Lu}V\Sigma_{Ld}^\dagger)=\exp(i(\Pi_{Lu}-\Pi_{Ld}+\sigma^2 \theta_{12}))
	=\exp(i\sigma_{12}(\theta_{12}+\frac{\kappa}{2(d_u^2-d_c^2)(d_d^2-d_s^2)}\varphi))\,,
\end{equation}
so that the CKM modes in this case enter like a shift of the Cabibbo angle in the scalar potential.

The interactions with the SM fermions read instead
\begin{equation}
\begin{aligned}
\frac{1}{2}{\overline u}_L 
\frac{d_d^2-d_s^2}{d_u^2-d_c^2}\frac{{d_u^2+d_c^2}}{\kappa}m_{c}( 1 -\frac{2 d_c^2}{d_u^2+d_c^2}) c_R \varphi
\simeq \frac{1}{2}\frac{d_d}{d_u\sqrt{d_u^2+d_d^2}}m_{c}{\overline u}_L  c_R \varphi 
\approx \frac{\sqrt2\lambda'_u m_u m_c}{\lambda_u M_u v} {\overline u}_L  c_R \varphi \\
\frac{1}{2}{\overline c}_L 
\frac{d_d^2-d_s^2}{d_u^2-d_c^2}\frac{{d_u^2+d_c^2}}{\kappa}m_{u} (\frac{2 d_u^2}{d_u^2+d_c^2}-1) u_R \varphi
\simeq \frac{1}{2}\frac{d_d}{d_u\sqrt{d_u^2+d_d^2}}m_{u}  {\overline c}_L u_R \varphi
\approx \frac{\sqrt2\lambda'_u m_u^2}{\lambda_u M_u v} {\overline c}_L  u_R \varphi
\end{aligned}
\end{equation}
and analogously for the down-type quarks. 
The most dangerous interaction is suppressed by $\frac{\lambda'_u m_u m_c}{\lambda_u M_u v}$ which, like for the diagonal modes,
provide an extra Yukawa suppression with respect to the Higgs coupling, which is already Yukawa suppressed.

In this case  the smallness of the coupling
guarantees no dangerous tree-level FC effects regardless of what the
masses of the radial modes may be, as long as they are above the
bounds from direct searches
 (and such bounds can be even quite loose because of the small couplings).

\subsection{The 3-flavors case: numerical}
In the 3-flavor case the formulae are lengthy and less intuitive, however one can still calculate
numerically the 3-fields vertices involving two SM fermions and one of the four CKM radial modes
(canonically normalized). Even without knowing the potential, and therefore the mass matrix of these
radial modes, one can estimate  their maximum FC contributions by assuming that the lightest
eigenmode couples to the vertices with the largest couplings. In this case  contributions
to $\Delta F=2$ operators of the form
$$ \frac{c}{m_{\pi}^2}(\overline q q)^2$$
are obtained, with
\begin{center}
\begin{tabular}{|c|cc|}
\hline 
&Re($c$)&Im($c$) \\ \hline
$(\overline s_R d_L)^2$ & $-7\cdot 10^{-18}$ & $-1\cdot10^{-20}$\\ \hline
$(\overline c_R u_L)^2$ & $-4\cdot 10^{-18}$ & $-3\cdot10^{-19}$\\ \hline
$(\overline b_R d_L)^2$ & $-7\cdot 10^{-17}$ & $-7\cdot10^{-22}$\\ \hline
$(\overline b_R s_L)^2$ & $-8\cdot 10^{-13}$ & $-9\cdot10^{-18}$\\ \hline
\end{tabular}
\end{center}
and with  $m_{\pi}$  the mass of the lightest CKM mode eigenstate. The contributions are so suppressed
that $m_\pi$ can be as light as 100~MeV without incurring into problems with flavor.
The quantitative results above nicely fit with what is observed in the two flavor case, and the couplings of
the canonically normalized CKM modes to the fermions are numerically compatible with
the short-hand formula
\begin{equation}
\pi_{\text{CKM}} \frac{\lambda'_{u,d} m_{q_i} m_{q_j}}{\lambda_{u,d} v M_{u,d}} \overline q_R^i q_L^j \,,
\end{equation}
which is similar to the flavor preserving one for the radial modes of the diagonal flavon fields.

\section{Oblique Corrections}
\label{appendixa}

In this appendix we derive the one loop expression for the $S$, $T$ and $U$ parameters in our model.

We use the standard definitions \cite{peskin},
\begin{equation}
\begin{aligned}
S&=-16 \pi \Pi_{3Y}'(0)\,,\\
T&=\frac {4\pi}{s_w^2 c_w^2 M_Z^2}\left[\Pi_{11}(0)-\Pi_{33}(0)\right]\,,\\
U&= 16 \pi \left[\Pi_{11}'(0)-\Pi_{33}'(0)\right].
\end{aligned}
\end{equation}
 
For simplicity we work in the limit where only the mixing of the top is important. 
From the couplings (\ref{Wcouplings}), (\ref{Zcouplings}) the contribution of the third 
generation to $T$ (obtained as the difference between the correlators in our model 
and their SM  values corresponding to $s_{u_{L3}}=0$) is given by,
\begin{multline}
T=\frac {3\pi}{s_w^2 c_w^2 M_Z^2}[ 2 s_{u_{L3}}^2 \Pi_{LL}(m_{t'},m_b,0)-2 s_{u_{L3}}^2  \Pi_{LL}(m_{t}, m_b,0)+
(1- c_{u_{L3}}^4)  \Pi_{LL}(m_t,m_t,0) \\- s_{u_{L3}}^4 \Pi_{LL}(m_{t'},m_{t'},0)- 2 s_{u_{L3}}^2 c_{u_{L3}}^2 \Pi_{LL}(m_{t'}, m_{t},0)]
\end{multline}
where we have introduced the self energies $\Pi_{LL}(m_1,m_2, q)$ with two left currents.
In the limit $m_b\to 0$ one finds,
\begin{equation}
T=\frac{3\,s_{u_{L3}}^2}{8 \pi\, s_w^2 c_w^2} \frac{m_t^2}{M_Z^2}\left [
c_{u_{L3}}^2\,\left(\frac{m_{t'}^2}{m_{t'}^2-m_{t}^2}\log \Bigl(\frac {m_{t'}^2}{m_t^2}\Bigr)-1 \right)
+\frac{s_{u_{L3}}^2}{2} \left(\frac{m_{t'}^2}{m_t^2}-1\right) \right]\,.
\label{T}
\end{equation}

Repeating the same steps for $S$ one obtains,
\begin{multline}
S= 4 \pi [s_{u_{L3}}^2 (3\,s_{u_{L3}}^2-2) \Pi_{LL}'(m_{t},m_{t},0)+s_{u_{L3}}^2 (3\,s_{u_{L3}}^2-4)\Pi_{LL}'(m_{t'},m_{t'},0) \\
+6 s_{u_{L3}}^2 c_{u_{L3}}^2 \Pi_{LL}'(m_t,m_{t'},0)+4 s_{u_{L3}}^2 \Pi_{LR}'(m_t,m_t,0)-4 s_{u_{L3}}^2 \Pi_{LR}'(m_{t'},m_{t'},0)]\,,
\end{multline}
which gives
\begin{multline}
S=\frac{s_{u_{L3}}^2}{6\pi}\left[
\left (3 c_{u_{L3}}^2\frac{(m_{t'}^2+m_t^2)(m_{t'}^4-4 m_{t'}^2 m_t^2+m_t^4)}{(m_{t'}^2-m_t^2)^3}-1\right)
\log \Bigl(\frac {m_{t'}^2}{m_t^2}\Bigr)\right.  \\
\qquad\left.-c_{u_{L3}}^2\frac{5 m_{t'}^4-22 m_{t'}^2 m_t^2+5m_t^4}{(m_{t'}^2-m_t^2)^2}\right]\,.
\label{S}
\end{multline}

For completeness the $U$ parameter is given by,
\begin{multline}
U=\frac{s_{u_{L3}}^2}{6\pi}\left[
-3\left ( c_{u_{L3}}^2\frac{(m_{t'}^2+m_t^2)(m_{t'}^4-4 m_{t'}^2 m_t^2+m_t^4)}{(m_{t'}^2-m_t^2)^3}-1\right)
\log \Bigl(\frac {m_{t'}^2}{m_t^2}\Bigr)\right.  \\
\qquad\left.+c_{u_{L3}}^2\frac{5 m_{t'}^4-22 m_{t'}^2 m_t^2+5m_t^4}{(m_{t'}^2-m_t^2)^2}\right]\,.
\end{multline} 

For our analysis we have used the recent analysis \cite{gfitter},
\begin{eqnarray}
\begin{array}{c} S=0.02\pm0.11 \\ T=0.05\pm0.12\\ U=0.07 \pm 0.12 \end{array}~~~~~~
{\rm with~correlation~matrix}~~~~~~ \left ( \begin{array}{ccc} 1 & 0.879 & -0.469\\ 0.879 & 1 &-0.716 \\-0.469 & -0.716 & 1 \end{array}\right).
\end{eqnarray}


\end{document}